\documentclass[structabstract]{aa}  
\usepackage{graphicx}
\usepackage{txfonts}
\usepackage{natbib}

\begin{document}

\title{Variable X-ray emission from the accretion shock \\ 
in the Classical T Tauri Star V2129~Oph}
\subtitle{}
\author{C.~Argiroffi \inst{1,2} \and E.~Flaccomio \inst{2} \and J.~Bouvier\inst{3} \and J.-F.~Donati\inst{4} \and K.~V.~Getman\inst{5} \and S.~G.~Gregory\inst{6,7} \and G.~A.~J.~Hussain\inst{8} \and M.~M.~Jardine\inst{9} \and M.~B.~Skelly\inst{4} \and F.~M.~Walter\inst{10}}
\institute{DSFA, Univ. di Palermo, Piazza del Parlamento 1, 90134 Palermo, Italy, \email{argi@astropa.unipa.it}
\and INAF - Osservatorio Astronomico di Palermo, Piazza del Parlamento 1, 90134 Palermo, Italy
\and UJF-Grenoble 1 / CNRS-INSU, Institut de Plan\'etologie et d'Astrophysique de Grenoble (IPAG) UMR 5274, Grenoble, F-38041, France
\and LATT–UMR 5572, CNRS \& Univ. de Toulouse, 14 Av. E. Belin, F–31400 Toulouse, France
\and Department of Astronomy \& Astrophysics, 525 Davey Laboratory, Pennsylvania State University, University Park, PA 16802, USA 
\and School of Physics, Univ. of Exeter, Stocker Road, Exeter EX4 4GL, UK
\and California Institute of Technology, MC 249-17, Pasadena, CA 91125 USA
\and ESO, Karl-Schwarzschild-Strasse 2, 85748 Garching bei M\"unchen, Germany
\and School of Physics and Astronomy, Univ. of St Andrews, St Andrews, Scotland KY16 9SS, UK
\and Department of Physics and Astronomy, Stony Brook University, Stony Brook, NY, 11794-3800, USA
}

\date{Received 14 December 2010; accepted 28 February 2011}
 
\abstract
{The soft X-ray emission from high density plasma observed in several CTTS is usually associated with the accretion process. However, it is still unclear whether this high density ``cool'' plasma is heated in the accretion shock, or if it is coronal plasma fed or modified by the accretion process.}
{We conducted a coordinated quasi-simultaneous optical and X-ray observing campaign of the CTTS V2129~Oph. In this paper we analyze Chandra grating spectrometer data and aim at correlating the observed X-ray emitting plasma components with the characteristics of the accretion process and of the stellar magnetic field constrained through simultaneous optical observations.}
{We analyze a 200\,ks Chandra/HETGS observation, subdivided into two 100\,ks segments, of the CTTS V2129~Oph. For the two observing segments, corresponding to two different phases within one stellar rotation, we measure the density of the cool plasma component and the emission measure distribution.}
{The X-ray emitting plasma covers a wide range of temperatures: from 2 up to 34\,MK. The cool plasma component of V2129~Oph ($T\approx3-4$\,MK) varies between the two segments of the Chandra observation: high density plasma ($\log N_{\rm e} = 12.1^{+0.6}_{-1.1}$) with high $EM$ at $\sim3-4$\,MK is present during the first observing segment; during the second segment this plasma component has lower $EM$ and lower density ($\log N_{\rm e} < 11.5$), although the statistical significance of these differences is marginal. Hotter plasma components, $T\ge 10\,MK$, show variability on short time scales ($\sim10$\,ks), typical of coronal plasma. A clear flare, detected during the first segment, could be located in a large coronal loop (half length $>3\,R_{\star}$).}
{Our observation provides further confirmation that the dense cool plasma at a few MK in CTTS is material heated in the accretion shock. The variability of this cool plasma component on V2129~Oph may be explained in terms of X-rays emitted in the accretion shock and seen with different viewing angles at the two rotational phases probed by our observation. In particular, during the first time interval direct view of the shock region is possible, while, during the second, the accretion funnel itself intersects the line of sight to the shock region, preventing us from observing the accretion-driven X-rays.}


\maketitle

\section{Introduction}

During the pre-main sequence phase low mass stars evolve from the embedded protostellar stage to the main sequence. In the first few million year of this transformation they undergo a key phase in which, while gravitationally contracting, they accrete material from a circumstellar disk \citep[see e.g. ][]{Bertout1989,KenyonHartmann1995}. During this phase low mass stars are named Classical T~Tauri stars (CTTS). When the accretion process ends, stars, still in the pre-main sequence but lacking the strong line emission characteristic of the previous phase, are classified as weak line T~Tauri stars (WTTS).

Recently it has been evidenced that the accretion process in CTTS is associated with the presence of high density plasma at temperature of a few MK radiating soft X-rays. Many fundamental aspects of this link are still debated. Moreover understanding the accretion process and its high-energy radiation is a key step in the understanding of the physics of pre-main sequence low mass stars and of their circumstellar environment, because accretion strongly affects several aspects of the central star itself: accretion determines stellar rotational velocities \citep{BouvierCabrit1993}, and affects coronal activity \citep{PreibischKim2005}; the accretion history in the early stages of stellar evolution might modify the stellar radius, and hence its luminosity, up to ages of a few Myr \citep{BaraffeChabrier2009}; in addition the accretion process, because of the intense ionizing/heating radiation produced (UV and X-rays), influences the physics and the chemistry of circumstellar disks where planets are forming and may even determine their lifetime \citep[e.g.][]{AlexanderClarke2004,ErcolanoClarke2009}.

The accretion process is a complex phenomenon in which a fundamental role is played by the stellar magnetic field: it truncates the inner part of the circumstellar disk and guides infalling material along its flux tubes toward the central star \citep{Konigl1991}. The magnetic field, interacting with the inner disk and channeling both inflowing (i.e. accreting) and outflowing material, mediates the angular momentum exchange between the star and the circumstellar material \citep{ShuNajita1994}.

The accreting material from the inner disk approaches the star in (almost) free fall. The impact with the dense stellar atmosphere produces shocks at the foot-points of the accretion streams. Most of the kinetic energy of the infalling material is thus converted into heat, generating a region at higher temperature with respect to the surrounding photosphere, a so-called hot spot. Evidence of such hot spots at $T\sim10\,000$\,K is provided by optical/UV photometry and spectroscopy of CTTS.

From a theoretical point of view accretion onto CTTS can produce significant amounts of X-rays, since the accreting material should impact onto the stellar surface with a high infall velocity ($v\sim300-500\,{\rm km\,s^{-1}}$), being thus heated in the accretion shocks up to temperatures of $\sim2-6$\,MK. Such hot material would radiate mostly in the soft X-ray band \citep[$0.1-1.0$\,keV, ][]{Gullbring1994}. The predicted X-ray luminosities in this band are large even for moderate mass accretion rates \citep[i.e. $L_{X}\sim10^{30}\,{\rm erg\,s^{-1}}$ for $\dot{M}\sim10^{-10}\,{\rm M_{\odot}\,yr^{-1}}$, ][]{SaccoArgiroffi2008}. The observed $L_{\rm X}$ from CTTS in the $0.1-1.0$\,keV band are, however, significantly lower than these predictions.

Observationally CTTS are X-ray bright when one considers a broader $0.1-10$\,keV band due to their very active coronae. In fact their intense magnetic fields, other than controlling accretion, are also responsible for the heating and confinement of large amounts of coronal plasma. Most of the X-ray emission from CTTS is actually due to very hot plasmas ($T\ge10$\,MK, much higher than that produced in accretion shocks) and is characterized by frequent flaring activity, clear evidence of coronal emission. Measurements of the density of the ``cool'' plasma ($T\sim3-4$\,MK) in CTTS are, however, at odds with a coronal origin of at least part of the X-ray emission: in 6 out of the 7 CTTS for which plasma densities have been measured \citep{KastnerHuenemoerder2002,StelzerSchmitt2004,SchmittRobrade2005,GuntherLiefke2006,ArgiroffiMaggio2007,GudelSkinner2007,RobradeSchmitt2007,HuenemoerderKastner2007} the cool plasma is significantly denser ($N_{\rm e}\ge10^{11}\,{\rm cm^{-3}}$), than ever observed in the same-temperature plasma of older late type stars \citep{TestaDrake2004} or in disk-bearing young stars without evidence of accretion \citep{KastnerHuenemoerder2004,GuntherMatt2010}. Such high density plasma cannot be interpreted in terms of steady coronal loops \citep{ArgiroffiMaggio2009}. These findings indicate that the high density plasma component in CTTS is associated with the accretion process. In particular it is suggested that this plasma is material heated in the accretion shock.

In addition to these studies focused on the plasma densities, interesting results have also been obtained by \citet{GudelTelleschi2007} that, investigating soft ($\sim0.5$\,keV) X-ray flux level of CTTS and WTTS, proved that CTTS have higher soft X-ray fluxes than WTTS. They suggest that this {\it soft excess} is due to shock heated plasma, or to magnetically confined plasma interacting with (and/or originating from) accretion funnels.

\citet{FlaccomioMicela2010}, by observing the NGC~2264 star forming region, found that time variability of the soft X-ray and optical luminosities are correlated in CTTS (and not correlated in WTTS). Such a correlation may be explained if X-rays and optical radiation are both produced in the accretion shock and are modulated by a time-variable accretion rate and/or by the stellar rotation. Supporting an idea previously proposed by \citet{GregoryWood2007}, \citeauthor{FlaccomioMicela2010}, however, favored a different scenario in which both the optical and soft X-ray variability are produced by a variable absorption, likely due to disk warps, as proposed by \citet{BouvierAlencar2007} for the CTTS AA~Tau.

All these results clearly indicate that the soft X-ray emission from high density plasma and the accretion process are somehow linked. Different scenarios have been depicted to explain this link, but many fundamental points are still discussed. The final evidence, clarifying whether the high density cool plasma component is material heated in the accretion shock \citep{KastnerHuenemoerder2002}, or whether it is coronal plasma somehow modified/affected by the accretion process \citep{GudelSkinner2007,BrickhouseCranmer2010}, is still lacking. Moreover a fundamental role is possibly played by a variable or non-uniform absorption produced by accretion streams and inner disk warps in CTTS.

Very recently a few high quality X-ray datasets of CTTS provided tighter observational constraints and evidenced important discrepancies with the predictions of current accretion-shock models. The high density cool plasma components of TW~Hya cannot be explained by the shock of a single accretion stream \citep{BrickhouseCranmer2010}, but models with multiple streams are required \citep{SaccoOrlando2010}. Moreover the shock-heated plasma model, applied to the observed X-rays from CTTS, provides significantly underestimated accretion rates with respect to measurements based on optical indicators \citep[a factor of 10 or even more, ][]{ArgiroffiMaggio2009,CurranArgiroffi2011}. Local absorption of X-rays by the surrounding stellar atmosphere \citep{Drake2005} or by the pre-shock infalling material could play a fundamental role in determining the amount of X-ray emission escaping from the shock region \citep{SaccoOrlando2010}.

The determination of the X-ray luminosity that emerges from the shock region, other than for interpreting the observations, is fundamental to: $i$) constrain the amount of high-energy radiation that heats and ionizes the circumstellar environment, and, $ii$) delineate the energy balance in the shock region.

In order to address the above issues, and hence contribute to the understanding of the accretion phenomenon as a whole, we have planned and performed a campaign of simultaneous multiwavelength observation of a CTTS. The immediate objective is to simultaneously constrain and correlate the characteristics of the accretion process, of the stellar magnetic field, and of the X-ray emitting plasma, including both the coronal and the accretion-related components \citep{GregoryFlaccomio2009}.  

We selected for this study V2129~Oph, a CTTS in the $\rho$~Ophiuchi star forming cloud and already one of the targets of the Magnetic Protostars and Planets (MaPP) project \citep{DonatiJardine2009}, whose main goal is studying the large-scale magnetic fields of CTTS using time-series of polarized Zeeman signatures.

In our study of V2129~Oph we benefit from several datasets, acquired simultaneously or quasi-simultaneously. V2129~Oph was monitored by optical spectroscopy and photometry from early May 2009 to mid July 2009. The observation set includes: high resolution X-ray spectroscopy obtained with the {\it Chandra} High Energy Transmission Grating Spectrometer (HETGS), optical spectroscopy both at low and high spectral resolution (SMARTS/RC Spectrograph, SMARTS/Echelle Spectrograph, ESO/HARPS), optical polarized spectroscopy (ESPaDOnS), and NIR and optical photometry (SMARTS/Andicam time series).

Results based on the SMARTS and ESPaDOnS data, including Zeeman-Doppler brightness and magnetic fields maps, are presented by \citet{DonatiBouvier2011} in a companion paper. In this paper we present the analysis of the X-ray {\it Chandra}/HETGS observation, taking
advantage, in the interpretation, of the results from the optical data. Magnetic field extrapolation, based on the Zeeman-Doppler maps, will be described in forthcoming papers and will allow us to derive a three-dimensional description of the open (wind-bearing) and closed (X-ray bright) field lines, and of the magnetic field structures through which accretion occurs \citep{JardineGregory2008}, for a more detailed comparison with the present X-ray observations.

We report the main properties of V2129~Oph in Sect.~\ref{v2129oph}. Sect.~\ref{dataanalysis} contains the analysis of the X-ray data, including short time scales variability (Sect.~\ref{shorttermvar}), and long time scales variability in which the analysis of high resolution X-ray spectra is possible (Sect.~\ref{highresspec}). Our main results are summarized in Sect.~\ref{results} and then discussed in Sect.~\ref{disc}.

\section{Properties of V2129~Oph}
\label{v2129oph}

V2129~Oph is a CTTS belonging to the $\rho$~Oph star forming region at a distance of 120\,pc. It is a K5 star, with $M=1.35\,{\rm M_{\sun}}$, $R_{\star}=2.1\,{\rm R_{\sun}}$, \citep[see ][ for details]{DonatiBouvier2011}. It has a lower mass companion, separated by
0.6\arcsec with a $K$-band flux ratio $\sim0.09$ with respect to the primary \citep{McCabeGhez2006}: using the evolutionary models of \citet{SiessDufour2000} the mass of the secondary is $\sim0.1\,{\rm M_{\sun}}$. Being located at a sky-projected distance of $\sim70$\,AU from V2129~Oph, we can reasonably assume that it does not significantly
affect its magnetosphere and accretion process. V2129~Oph has strong and variable H${\alpha}$ emission \citep[$EW\sim10-14\,\AA$, ][]{BouvierAppenzeller1992}. We selected V2129~Oph for this campaign because it is bright both in the optical and the X-ray band, suffers moderate interstellar absorption \citep[$A_{\rm V}\approx0.6$,][]{DonatiJardine2007}, and because, thanks to its rotational period \citep[$P_{\rm rot}=6.53$\,d, ][]{ShevchenkoHerbst1998,GrankinBouvier2008} and inclination angle \citep[$i\approx60\,\deg$,][]{DonatiBouvier2011}, it can be studied with techniques that exploit the rotational modulation of surface features. V2129~Oph is the first CTTS for which Zeeman-Doppler maps of the surface magnetic fields were obtained \citep{DonatiJardine2007} from time series of spectropolarimetric observations performed in 2005. Following this study \citet{JardineGregory2008} extrapolated the surface magnetic field maps and inferred the characteristics of the accretion funnels and of the coronal plasma of V2129~Oph at the time of the 2005 campaign. No coordinated X-ray observations were performed at that time, and indeed, previous to our observation, V2129~Oph had been only observed in X-rays by {\it ROSAT} with low sensitivity and spectral resolution.

During our 2009 observing campaign we acquired a new ESPaDOnS time series during  $1-14$ July, almost simultaneously with the {\it Chandra} observation (27-29 June 2009, see Sect.~\ref{dataanalysis}). The derived properties of V2129~Oph at the time of the {\it Chandra} observation are presented by \citet{DonatiBouvier2011}. The presence of a hot spot
covering $\sim$2.5\% of the stellar surface and located at high latitude ($\sim60\,\deg$) was indicated by Zeeman-Doppler maps of the \ion{Ca}{ii} IRT excess emission. This excess emission is powered by the accretion process, and therefore indicates the location of the footpoint of the main accretion stream in the visible part of the stellar surface. \citeauthor{DonatiBouvier2011} find that the radial component of the magnetic field has its peak intensity ($\sim4\,{\rm kG}$) in this same high latitude region, thus supporting its identification as the base of the accretion funnel, i.e. the magnetic flux tube that channels the accreting material. This accretion funnel is likely trailing the hot spot (i.e. its base), during the stellar rotation, as indicated by the periodic red-shifted absorption observed in the profiles of the Balmer lines. The intensity of the magnetic field suggests that the inner disk is truncated at a distance of $7.2\,R_{\star}$. The accretion rate of V2129~Oph appears to be quite stable during the observing campaign (apparent variations within a factor 2) with an average value of $\log \dot{M}=-9.2$, with $\dot{M}$ in units of ${\rm M_{\sun}\,yr^{-1}}$.

The upper part of Fig.~\ref{fig:lc} shows surface maps of the excess emission and of the radial component of the magnetic field from \citet{DonatiBouvier2011}. The maps are shown as viewed from the observer at rotational phases corresponding to the beginning and to the end of each of the two {\em Chandra} exposures comprising our X-ray observation.

\section{Chandra observation and data analysis}
\label{dataanalysis}

\begin{figure*}[t]
\centering
\includegraphics[width=18cm]{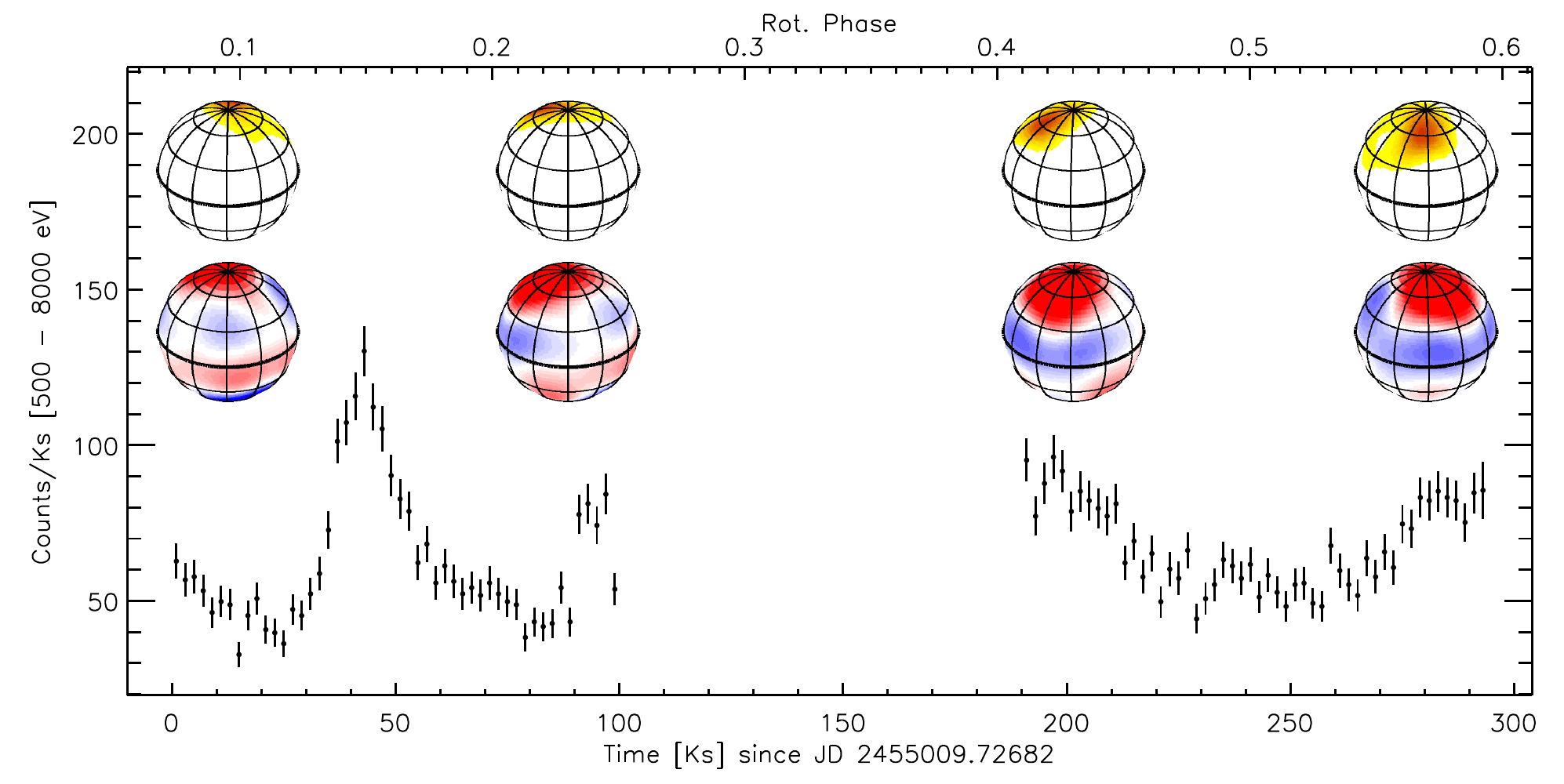}
\caption{{\it Chandra} light curve in the 0.5-10.0\,keV band combining events from the zero-order image and the HEG and MEG first order spectra, with 2\,ks time bin. Shown above the light curve are images of the excess emission due to accretion (upper row) and of the radial component of the magnetic field (lower row) as reconstructed with quasi-simultaneous Zeeman-Doppler imaging by \citet{DonatiBouvier2011}. Colors scales are as in \citeauthor{DonatiBouvier2011} (red and blue in the lower row indicate positive and negative B-fields, respectively). The images refer to rotational phases at the beginning and at the end of each of the two X-ray exposures.}
\label{fig:lc}
\end{figure*}

V2129~Oph was observed by {\it Chandra}/HETGS on 27-29 June 2009, for a total exposure of 200.4\,ks. The observation, performed during two consecutive orbits, was divided into two segments (obs. id. 9943 and 9944), with durations of 98.9 and 101.7\,ks, and separated by 88.6\,ks. The rotational phases of the surface maps of excess \ion{Ca}{ii} emission and radial magnetic field \citep{DonatiBouvier2011}, corresponding to starts and stops of the {\it Chandra} exposures, are shown in the upper part of Fig.~\ref{fig:lc}. Rotational phases referenced throughout this paper are referred to the ephemeris defined in \citet{DonatiJardine2007}.

\subsection{Data preparation}
\label{dataproc}

The HETGS on board of {\em Chandra} \citep{CanizaresDavis2005} is composed by two gratings, the High Energy and the Medium Energy Gratings (HEG and MEG). The HEG and MEG $1^{\rm st}$ order spectra cover the $1-15$ and $2-30$\,\AA~ranges, with resolutions of
0.012 and 0.023\,\AA (FWHM), respectively.

We reprocessed the {\it Chandra}/HETGS data using CIAO 4.2 (CALDB version 4.3.1) starting from the level 1 event file and following the standard processing threads described in the observatory website. For the analysis we used both the dispersed HEG and MEG photons, and the $0^{\rm th}$ order image. For the $0^{\rm th}$ order image, source photons were extracted from a circular region centered on the source position (with radius of 5\,px or $\sim2.5$\,\arcsec, expected to encircle $>95$\% of the source photons), while the background was taken from a concentric annulus (with 10 and 15\,px as inner and outer radii). HEG and MEG dispersed spectra were extracted from the standard rectangular regions aligned with the dispersion directions.

\begin{figure*}[t]
\centering
\includegraphics[width=18cm]{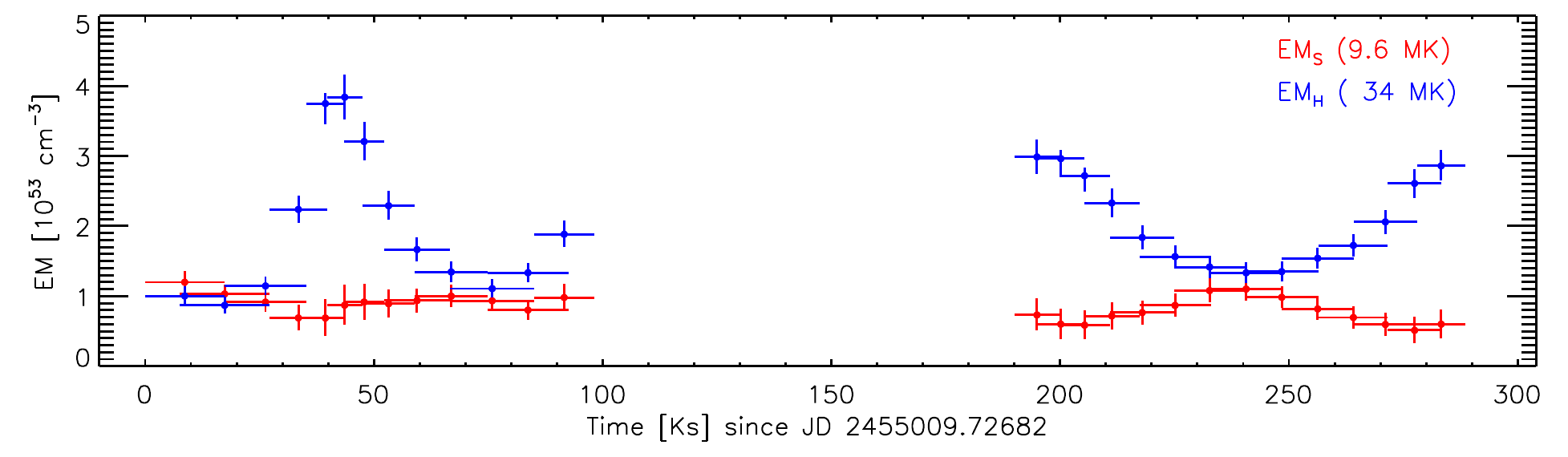}
\caption{Emission measure vs. time. Red and blue points refer to the EM of the cool and hot components, respectively. See Sect.~\ref{timeresspec} for details.}
\label{fig:2Tem_vs_time}
\end{figure*}

\subsection{Zero order image and CCD-resolution spectrum}
\label{0ordermean}

We inspected the shape of the point spread function (PSF) of the $0^{\rm th}$ order image in order to assess the possible contribution to the detected X-rays from the low-mass companion to V2129~Oph. With a 0.6\arcsec separation, and considering the {\it Chandra} spatial resolution, the companion might be marginally resolved if X-ray bright. The PSF, however, did not show evidence of asymmetry. We will therefore assume that the observed X-rays are all from V2129~Oph. This is in agreement with the expected low X-ray luminosity of the secondary that, given the mass-$L_{\rm X}$ relation of \citet{PreibischKim2005}, is $\sim8\times10^{29}\,{\rm erg\,s^{-1}}$, about 3\% of the observed $L_{\rm X}$.

We then analyzed the CCD ACIS spectrum obtained from the undispersed $0^{\rm th}$ order image with the main goal to constrain the amount of absorption affecting X-rays from V2129~Oph. Absorption, parametrized as usual by the equivalent hydrogen column density $N_{\rm H}$, alters the shape of the X-ray spectrum mostly at energy below 1\,keV (for modest values of $N_{\rm H}$) and, in this energy range, the $0^{\rm th}$ order spectrum has a higher signal to noise ratio than HEG and MEG spectra.

We considered the $0^{\rm th}$ order spectrum integrated over the whole {\it Chandra} observing time (we found no evidence of $N_{\rm H}$ variations within the {\it Chandra} observation, see Sect.~\ref{shorttermvar}). The spectrum, binned between 0.5 and 10 keV so to have a signal to noise ratio of 10.0 per bin, was analyzed using XSPEC. A two-temperature thermal model ({\sc apec}, in XSPEC), subject to interstellar absorption ({\sc wabs}) was found to adequately fit the observed spectrum. Elemental abundances were kept frozen to the reference values obtained by \citet{MaggioFlaccomio2007} from the X-ray emission of young stars of the Orion Nebula Cluster (ONC). We evaluated the statistical quality of the fits from the null probability implied by the $\chi^2$ and the number of degrees of freedom. The parameters of the emission model are: $T_{1}=9.6^{+1.5}_{-1.0}$, $T_{2}=34^{+5}_{-4}$\,MK, $\log EM_{1}=52.94\pm0.11$, and $\log EM_{2}=53.21^{+0.05}_{-0.07}$ (with $EM$ in units of ${\rm cm^{-3}}$). The best fit value for the hydrogen column density, $N_{\rm H}$, is $(8^{+6}_{-4})\times10^{20}\,{\rm cm^{-2}}$, fully compatible with the value estimated from $A_{V}$ using a standard conversion relation: $N_{\rm H}=1.6~A_{\rm V}\times 10^{21}$ \citep{VuongMontmerle2003}. Finally we note that the best fit $N_{\rm H}$ value does not change significantly if the $0^{\rm th}$ order spectrum is fit adopting the abundances that result from our analysis of the dispersed spectrum (Sect.~\ref{emd}).

\subsection{Short term variability}
\label{shorttermvar}

In order to investigate time-variability on short timescales we combined the $0^{\rm th}$ order undispersed photons with the $1^{\rm st}$ order HEG and MEG spectra, collected during selected time-intervals. Combining these three datasets allows us to increase the signal to noise ratio and hence to explore time scales down to a few hours.

\subsubsection{X-ray light curve}

We first constructed a combined background-subtracted light curve from the arrival times of photons in {\it source} and {\it background} regions of both the $0^{\rm th}$ order image and the HEG and MEG dispersed spectra. For the 0$^{\rm th}$ order image, source and background photons were extracted as indicated in Sect.~\ref{dataproc}. For the HEG and MEG spectra, in order to increase the signal to noise ratio, we selected {\it non standard} regions: $1^{\rm st}$ order source events were extracted within 2\arcsec-wide strips along the direction of dispersion; background events from strips, extending from 7.2 to 14.4\arcsec perpendicular to dispersion direction, on each side of each spectrum. Considering the two observing segments, we collected, after background subtraction, a total of 5852, 2034.5, and 5341.25 photons (in the E=0.5-10\,keV band) from the $0^{\rm th}$ order image, the HEG spectrum, and the MEG spectrum, respectively.

Figure~\ref{fig:lc} shows the $0^{\rm th}+1^{\rm st}$ order light curve of our two observing segments, using 2\,ks binning. The time elapsed since the beginning of the observation and the rotational phase of V2129\,Oph are given in the lower and upper x-axis, respectively.

The light curve shows a clear flare toward the middle of the first segment, lasting $\sim$10 hours and with peak count-rate $\sim3$ times higher than the pre-flare level, and what appears as a smaller flare toward the end of the same segment. During the second segment a seemingly more gradual variability is observed with a slow decay in the first $\sim$35\,ksec, and a similarly slow rise in the last $\sim$35\,ksec.

\subsubsection{Time resolved spectroscopy}
\label{timeresspec}

A better insight on the physical origin of the observed variability might come from a time-resolved spectral characterization of the X-ray emission. To this end we have extracted spectra in time intervals containing a total of 900 counts in the combined $0^{\rm th}$ order and $1^{\rm st}$ order (HEG+MEG) light curve. We {\it oversampled} the light curve so that each $i^{\rm th}$ interval overlaps with the ($i\pm$1)$^{\rm th}$ ones but is fully independent from the ($i\pm$2)$^{\rm th}$ intervals. For each interval three spectra were extracted: the $0^{\rm th}$ order ACIS spectrum, and the two $1^{\rm st}$ order HEG and MEG spectra. These spectra were binned between 0.5 and 10 keV so to have a signal to noise ratio of 3.0 per bin. Simultaneous spectral fits to the three spectra\footnote{Note that for the analysis of the time-average spectrum we used the $0^{\rm th}$ order data only. This was justified by the higher signal to noise ratio of the $0^{\rm th}$ order spectrum at low energies and the need to avoid cross-calibration issues. For the lower signal spectra discussed here the statistical uncertainties are always larger than the systematic ones and cross-calibration issues are therefore less important.} in the $0.5-10$\,keV band were then performed with XSPEC, using both the $\chi^2$-statistic and the Cash-statistic.

We adopted 1- and 2- temperatures thermal models ({\sc apec}, in XSPEC), including the effect of interstellar absorption ({\sc wabs}), with plasma abundances fixed to the values of \citet{MaggioFlaccomio2007}. Fits were performed both leaving the absorption, i.e. the hydrogen column density $N_{\rm H}$, as a free parameter and fixing $N_{\rm H}$ to $8\times10^{20}\,{\rm cm^{-2}}$, the value obtained in Sect.~\ref{0ordermean} for the time averaged $0^{\rm th}$ order spectrum. The statistical quality of the fits was evaluated by considering the null probability associated to the $\chi^2$ and the number of degrees of freedom. Two thermal components were generally required to obtain acceptable fits, especially when fixing the $N_{\rm H}$. No evidence of variability of the absorption was found and we thus favored models with fixed $N_{\rm H}$. Although two temperatures were always required for an acceptable fit, no evidence for a significant evolution of the temperatures was found (the only exception being the time interval corresponding to the rise phase of the flare where a different modeling approach indicates a higher temperature for the hot component, see Sect.~\ref{flare}). Statistically acceptable fits could be obtained by fixing the temperatures to those obtained from the average spectrum (Sect.~\ref{0ordermean}). This simply indicates that the Emission Measure Distribution ($EMD$) of the plasma, while varying significantly in time, can at all times be schematically represented by two dominant thermal components at fixed temperatures and that no significant additional component can be discerned with the limited resolution and signal of the spectra in individual time intervals\footnote{We have also repeated the analysis with a coarser time-binning, i.e. adopting intervals that contain 2000 counts, instead of 900, with analogous results.}. Fixing the $N_{\rm H}$ and the two temperatures, the only free parameters left in the fits are the emission measures, $EM$, of the two components. Figure~\ref{fig:2Tem_vs_time} shows the time evolution of the cool and hot $EM$. A hardening of the spectrum at times of high count-rates is observed (and is also evidenced by a simpler hardness ratio analysis) both during the flares (see Sect.~\ref{flare}) in the first observing segment and in the more gradual evolution in the second segment.

From the above analysis, the absorption-corrected X-ray luminosity of V2129~Oph ranges, during our {\it Chandra} observation, from 2.1 to $5.5\times10^{30}\,{\rm erg\,s^{-1}}$ in the $0.5-10.0$\,keV band.

\begin{figure}
\centering
\includegraphics[width=9cm]{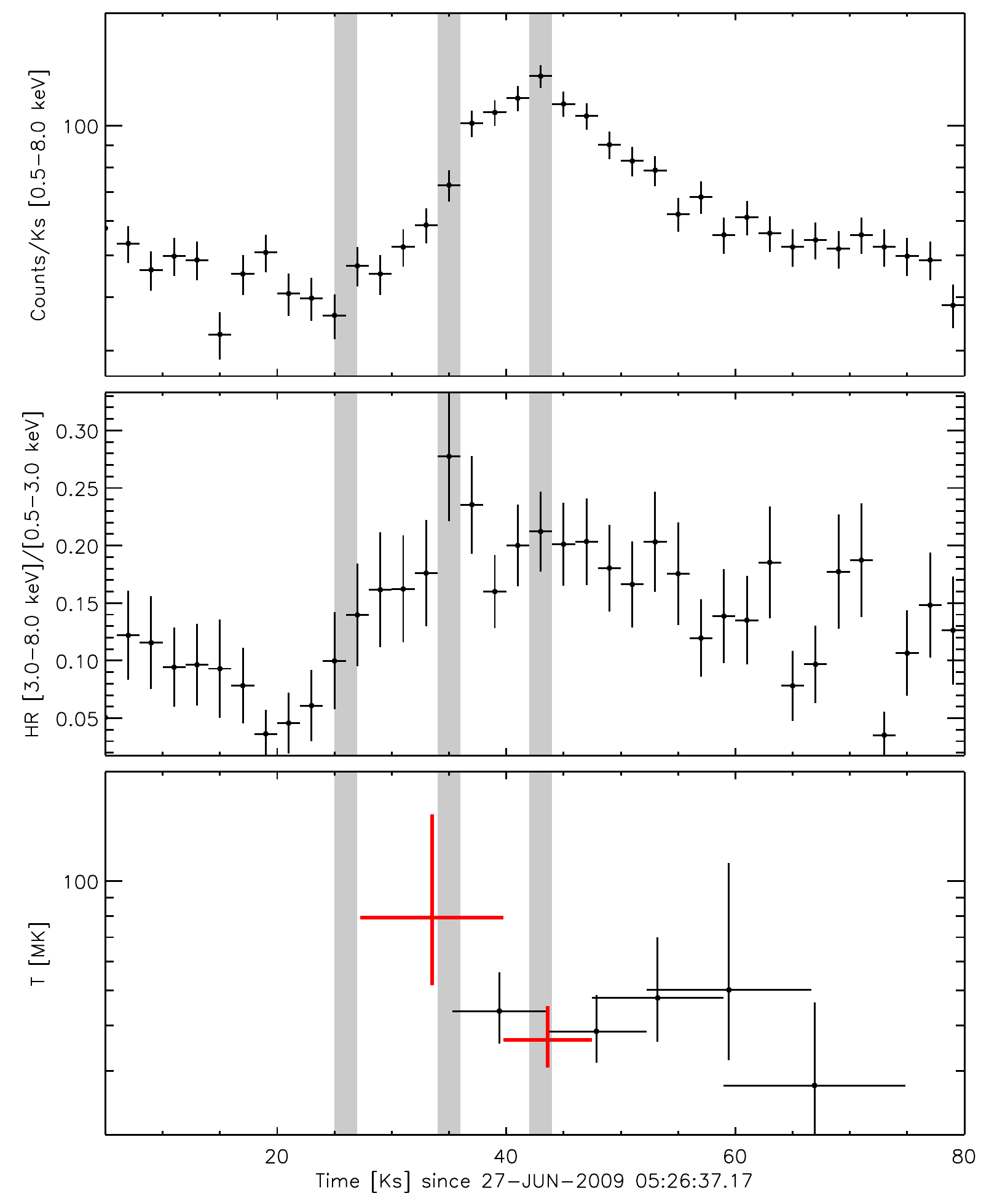}
\caption{{\it Upper panel}: {\it Chandra} X-ray light curve (the same shown in Fig.~\ref{fig:lc}) of the flare. {\it Middle panel}: Hardness ratio ($HR$), defined as the ratio between events in the $3.0-8.0$\,keV band and events in the $0.5-3.0$\,keV, vs. time. {\it Lower panel}: Temperature of the hot plasma component during flare evolution. Red points mark the maximum temperature and the temperature measured at the time in which the count rate is maximum.}
\label{fig:flare}
\end{figure}

\subsubsection{Flare analysis}
\label{flare}

Flares offer the opportunity to constrain the spatial distribution of coronal plasma. In particular, the length of the magnetic loops where flares occur can be estimated from the evolution of observable plasma properties during both the rise and the decay phases \citep{RealeBetta1997,Reale2007}. We have thus analyzed in detail the large flare detected during the first observing segment. 

In order to use the diagnostic based on the decay phase, the heating occurring during the decay must be characterized. This is done by studying the time-evolution of the plasma temperature and density. In our case, however, a time-resolved spectral analysis yielded no evidence of a significant temperature evolution during the decay phase, preventing the application of this method.

An estimate of the loop length could instead be obtained from the rise phase of the flare \citep{Reale2007}. The quantities used to derive the loop length are: the time $t_{0}$ at which the loop temperature reaches its maximum, the corresponding plasma temperature
$T_{0}$, the time at which the plasma density is maximum, $t_{\rm M}$,
and the corresponding plasma temperature, $T_{\rm M}$. Both $t_{0}$ and
$t_{\rm M}$ are relative to the beginning of the flare and the temperatures refer to the apex of the loop. Two independent estimates of the loop half length $L$ are then given by \citep[eq.~12 and 13 of][]{Reale2007}:

\begin{equation}
\label{length}
L \approx 3\,\Psi^2\,T_{0}^{1/2}\,t_{\rm M} \\
L \approx 2.5\,\frac{\Psi^2}{\ln \Psi}\,T_{0}^{1/2}\,(t_{\rm M}-t_{0})
\end{equation}

\noindent
where $\Psi$ is $T_{0}/T_{\rm M}$, $L$ is in units of $10^{9}$\,cm, $T_{0}$ is in units of $10^{7}$\,K, $t_{0}$ and $t_{\rm M}$ are in ks.

We applied these diagnostics to the flare of V2129~Oph by analyzing the $0^{\rm th}+1^{\rm st}$ order HEG and MEG events. The times $t_{0}$ and $t_{\rm M}$ were estimated by inspecting the time-evolution of the count-rate and hardness ratio ($HR$, the ratio between the count rate in the hard and in the soft band). The time of maximum density corresponds to the time of maximum count rate: this latter is indeed approximately proportional to the $EM$ of the plasma in the flaring loop and therefore, assuming that the loop volume is constant, a proxy of the plasma density. The time of maximum temperature corresponds
to the time of maximum $HR$, since the $HR$ increases for increasing temperature of the flaring plasma.

The upper panel of Fig.~\ref{fig:flare} shows the observed count rate in the $0.5-8.0$\,keV X-ray band during the flare. We estimate that the flare starts $26\pm1$\,ks after the beginning of the {\it Chandra} observation, and that the density peak occurs at $t_{\rm
M}=43\pm1$\,ks. The middle panel of Fig.~\ref{fig:flare} shows the evolution of the $HR$, evaluated adopting $0.5-3.0$ and $3.0-8.0$\,keV as the soft and hard bands so to maximize the sensitivity to high temperatures. The maximum temperature occurs at $t_{0}=35\pm1$\,ks. The beginning of the flare and the times of maximum temperature and density are indicated in Fig.~\ref{fig:flare} with vertical gray bands.

We determined the temperatures $T_{0}$ and $T_{\rm M}$ through time resolved spectral fittings. The adopted time intervals and the resulting extracted spectra are the same as defined in Sect.~\ref{timeresspec}. Being interested in the temperatures of the flaring plasma we must subtract the X-ray emission from the ``quiescent'' corona. We do so by assuming that this latter remains constant during the flare and is well represented by the X-ray spectrum registered just before the beginning of the flare (in particular, with
reference to Fig. \ref{fig:2Tem_vs_time}, during the second interval centered at $t\sim$18\,ks). We thus first derived a spectral model for the quiescent emission, and then fitted the spectra of the ``flaring'' time-intervals assuming a spectral model composed by the sum of the quiescent spectrum (with fixed parameters) and an optically-thin isothermal plasma component ({\sc apec}, in XSPEC) subject to the same interstellar extinction as derived from the time-averages spectral analysis ($N_{\rm H}=8\times10^{20}\,{\rm cm^{-2}}$). This latter component, describing the flaring plasma, is characterized by two fit-parameters, the temperature $T$ and the emission measure $EM$. The resulting best-fit temperatures are shown in the lower panel of Fig.~\ref{fig:flare}, with thicker red symbols marking the values of the maximum temperature, $80^{+70}_{-30}$\,MK, and of the temperature at the time of maximum density, $37^{+9}_{-6}$\,MK. Note that these values are averages over the real temperature distribution of the flaring plasma, and do not directly correspond to the $T_{0}$ and $T_{\rm M}$ used in eq.~\ref{length} that are the temperatures at the loop apex. $T_{0}$ and $T_{\rm M}$ can however be obtained from the measured temperatures with the transformation formulas reported in \citet{Reale2007}, yielding $T_{0}=210^{+240}_{-80}$\,MK and $T_{\rm M}=81^{+24}_{-16}$\,MK.

\begin{figure*}
\centering
\includegraphics[width=18cm]{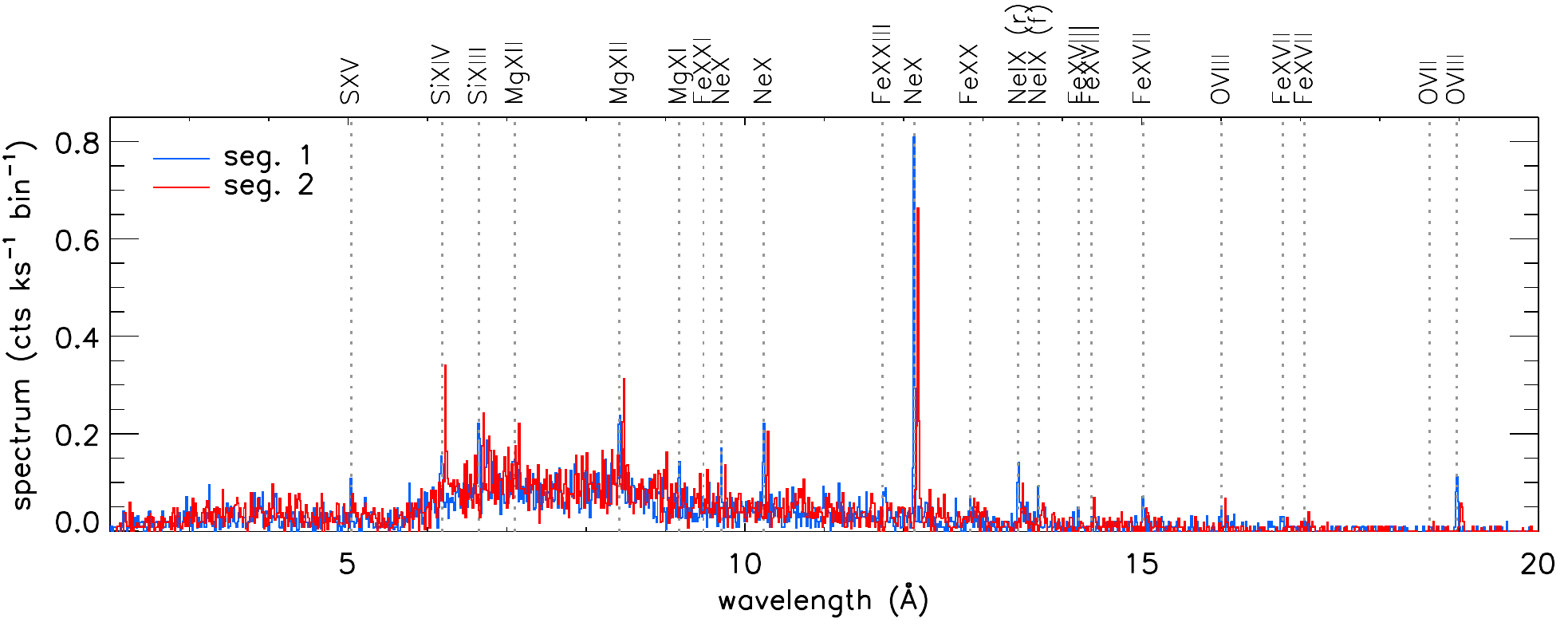}
\caption{V2129~Oph X-ray spectra observed with {\em Chandra} (HEG+MEG $1^{\rm st}$ orders) during the two observing segments.  For clarity reasons both spectra are rebinned by a factor of 3 (bin size~$=0.015$\,\AA), and the spectrum of seg.~2 is shifted toward longer wavelengths by $0.05\,\AA$.}
\label{fig:spec}
\end{figure*}

\begin{table*}
\caption{Measured fluxes of emission lines and continuum intervals in the X-ray spectrum of V2129~Oph.}
\label{tab:fluxes}
\small
\begin{center}
\begin{tabular}{r@{$\div$}llcr@{$\;\pm\;$}lr@{$\;\pm\;$}lr@{$\;\pm\;$}l}
\hline\hline
 \multicolumn{2}{c}{ } & & & \multicolumn{2}{c}{seg. 1} & \multicolumn{2}{c}{seg. 2} & \multicolumn{2}{c}{seg. 1+2} \\
\multicolumn{2}{c}{$\lambda^{a}$} & Ion & $\log T_{\rm max}^{b}$ & \multicolumn{2}{c}{Flux$^{c}$} & \multicolumn{2}{c}{Flux$^{c}$} & \multicolumn{2}{c}{Flux$^{c}$} \\
\hline
\multicolumn{10}{c}{\it Lines} \\
\hline
\multicolumn{2}{r}{ 5.04\hspace{5mm} } &                                                            \ion{S}{xv} &  7.20 &         1.2 &         0.7 &         0.2 &         0.6 &         0.7 &         0.4 \\
\multicolumn{2}{r}{ 6.18\hspace{5mm} } &                                            \ion{Si}{xiv} \ion{Si}{xiv} &  7.20 &         0.8 &         0.3 &         1.3 &         0.4 &         1.1 &         0.3 \\
\multicolumn{2}{r}{ 6.65\hspace{5mm} } &                                                         \ion{Si}{xiii} &  7.00 &         0.9 &         0.4 &         1.0 &         0.4 &         1.0 &         0.3 \\
\multicolumn{2}{r}{ 6.69\hspace{5mm} } &                                          \ion{Si}{xiii} \ion{Si}{xiii} &  6.95 &         0.0 &         0.3 &         0.1 &         0.3 &         0.0 &         0.2 \\
\multicolumn{2}{r}{ 6.74\hspace{5mm} } &                                                         \ion{Si}{xiii} &  7.00 &         0.9 &         0.4 &         1.0 &         0.3 &         0.9 &         0.2 \\
\multicolumn{2}{r}{ 7.11\hspace{5mm} } &                                            \ion{Mg}{xii} \ion{Mg}{xii} &  7.00 &         0.3 &         0.3 &         0.7 &         0.3 &         0.5 &         0.2 \\
\multicolumn{2}{r}{ 8.42\hspace{5mm} } &                                            \ion{Mg}{xii} \ion{Mg}{xii} &  7.00 &         1.1 &         0.4 &         1.5 &         0.4 &         1.3 &         0.3 \\
\multicolumn{2}{r}{ 9.17\hspace{5mm} } &                                                           \ion{Mg}{xi} &  6.80 &         0.9 &         0.5 &         0.7 &         0.5 &         0.8 &         0.3 \\
\multicolumn{2}{r}{ 9.23\hspace{5mm} } &                                                           \ion{Mg}{xi} &  6.80 &         0.5 &         0.5 &         0.0 &         0.3 &         0.1 &         0.3 \\
\multicolumn{2}{r}{ 9.31\hspace{5mm} } &                                                           \ion{Mg}{xi} &  6.80 &         0.0 &         0.4 &         0.3 &         0.4 &         0.2 &         0.3 \\
\multicolumn{2}{r}{ 9.48\hspace{5mm} } &                                  \ion{Ne}{x} \ion{Ne}{x} \ion{Fe}{xxi} &  7.00 &         0.2 &         0.4 &         0.8 &         0.5 &         0.5 &         0.3 \\
\multicolumn{2}{r}{ 9.71\hspace{5mm} } &                                                \ion{Ne}{x} \ion{Ne}{x} &  6.80 &         1.2 &         0.6 &         0.9 &         0.5 &         1.0 &         0.4 \\
\multicolumn{2}{r}{10.24\hspace{5mm} } &                                                \ion{Ne}{x} \ion{Ne}{x} &  6.80 &         2.5 &         0.7 &         2.7 &         0.7 &         2.6 &         0.5 \\
\multicolumn{2}{r}{11.74\hspace{5mm} } &                                                        \ion{Fe}{xxiii} &  7.20 &         0.0 &         0.6 &         0.8 &         0.7 &         0.4 &         0.4 \\
\multicolumn{2}{r}{11.77\hspace{5mm} } &                                                         \ion{Fe}{xxii} &  7.10 &         1.6 &         0.9 &         0.6 &         0.7 &         1.1 &         0.5 \\
\multicolumn{2}{r}{12.13\hspace{5mm} } &                                                \ion{Ne}{x} \ion{Ne}{x} &  6.80 &        19.9 &         2.4 &        15.7 &         2.1 &        17.8 &         1.5 \\
\multicolumn{2}{r}{12.26\hspace{5mm} } &                                                         \ion{Fe}{xvii} &  6.80 &         0.5 &         0.8 &         0.9 &         0.8 &         0.7 &         0.5 \\
\multicolumn{2}{r}{12.28\hspace{5mm} } &                                                          \ion{Fe}{xxi} &  7.00 &         1.1 &         0.9 &         1.6 &         0.9 &         1.4 &         0.6 \\
\multicolumn{2}{r}{12.83\hspace{5mm} } &                                              \ion{Fe}{xx} \ion{Fe}{xx} &  7.00 &         1.3 &         1.2 &         1.4 &         1.2 &         1.3 &         0.8 \\
\multicolumn{2}{r}{13.45\hspace{5mm} } &                                                           \ion{Ne}{ix} &  6.60 &         8.1 &         2.4 &         3.6 &         1.8 &         5.8 &         1.4 \\
\multicolumn{2}{r}{13.55\hspace{5mm} } &                                                           \ion{Ne}{ix} &  6.55 &         2.3 &         1.7 &         0.3 &         1.2 &         1.3 &         0.9 \\
\multicolumn{2}{r}{13.70\hspace{5mm} } &                                                           \ion{Ne}{ix} &  6.60 &         2.8 &         1.8 &         2.8 &         1.7 &         2.8 &         1.1 \\
\multicolumn{2}{r}{14.20\hspace{5mm} } &                                        \ion{Fe}{xviii} \ion{Fe}{xviii} &  6.90 &         2.2 &         2.0 &         0.6 &         1.6 &         1.4 &         1.1 \\
\multicolumn{2}{r}{14.37\hspace{5mm} } &                        \ion{Fe}{xviii} \ion{Fe}{xviii} \ion{Fe}{xviii} &  6.90 &         1.2 &         1.6 &         1.9 &         1.7 &         1.6 &         1.0 \\
\multicolumn{2}{r}{15.02\hspace{5mm} } &                                                         \ion{Fe}{xvii} &  6.75 &         5.5 &         2.8 &         3.7 &         2.4 &         4.6 &         1.7 \\
\multicolumn{2}{r}{16.01\hspace{5mm} } &                            \ion{Fe}{xviii} \ion{O}{viii} \ion{O}{viii} &  6.50 &         3.9 &         3.1 &         7.0 &         3.5 &         5.5 &         2.1 \\
\multicolumn{2}{r}{16.78\hspace{5mm} } &                                                         \ion{Fe}{xvii} &  6.70 &         4.8 &         3.7 &         2.3 &         3.1 &         3.5 &         2.1 \\
\multicolumn{2}{r}{17.05\hspace{5mm} } &                                                         \ion{Fe}{xvii} &  6.70 &         3.0 &         3.7 &         5.0 &         4.1 &         4.0 &         2.4 \\
\multicolumn{2}{r}{17.10\hspace{5mm} } &                                                         \ion{Fe}{xvii} &  6.70 &         1.7 &         3.4 &         3.8 &         3.9 &         2.8 &         2.2 \\
\multicolumn{2}{r}{18.63\hspace{5mm} } &                                                           \ion{O}{vii} &  6.35 &         1.5 &         4.5 &         3.4 &         5.0 &         2.4 &         2.8 \\
\multicolumn{2}{r}{18.97\hspace{5mm} } &                                            \ion{O}{viii} \ion{O}{viii} &  6.50 &        45.3 &        13.2 &        18.3 &         9.3 &        31.6 &         7.5 \\
\hline
\multicolumn{10}{c}{\it Continuum} \\
\hline
            $[ 2.01 $ & $  2.51]$ &                                                                        &       &         4.9 &         1.0 &         6.6 &         1.1 &         5.8 &         0.7 \\
            $[ 2.51 $ & $  2.99]$ &                                                                        &       &         5.2 &         0.8 &         5.7 &         0.9 &         5.5 &         0.6 \\
            $[ 3.38 $ & $  3.64]$ &                                                                        &       &         3.8 &         0.7 &         5.7 &         0.8 &         4.8 &         0.5 \\
            $[ 3.78 $ & $  3.88]$ &                                                                        &       &         2.7 &         0.6 &         2.8 &         0.6 &         2.7 &         0.4 \\
            $[ 4.33 $ & $  4.67]$ &                                                                        &       &         5.6 &         1.1 &         8.7 &         1.3 &         7.2 &         0.8 \\
            $[ 6.29 $ & $  6.53]$ &                                                                        &       &         5.4 &         0.6 &         6.5 &         0.6 &         5.9 &         0.4 \\
\hline
\end{tabular}
\end{center}
$^a$~Wavelengths (\AA).
$^b$~Temperature (K) of maximum emissivity.
$^c$~Observed fluxes (${\rm 10^{-6}\,ph\,s^{-1}\,cm^{-2}}$) with uncertainties at the 68\% confidence level.
\normalsize
\end{table*}
\normalsize

From the above estimates of $t_{0}$, $t_{\rm M}$, $T_{0}$, and $T_{\rm M}$, we finally derive two estimates for the length of the flaring loop $L$ with the formulas in eq.~\ref{length}. In order to estimate uncertainties on $L$, we performed Monte Carlo simulations\footnote{This procedure was preferred to simple error propagation which requires uncertainties to be small, since this is not the case for our two temperatures.} of the distribution of the input parameters $t_{0}$, $t_{\rm M}$, $T_{0}$, and $T_{\rm M}$, and inferred the $L$ quantile corresponding to $\pm1\sigma$. The uncertainties on $L$ are dominated by those on $T_{0}$ and $T_{\rm M}$. In particular the poorly constrained upper limit on $T_{0}$ yields a poorly constrained upper limit also on $L$. Both expression for $L$ provide very similar results, supporting the validity of the analysis. The best values for $L$ are $10-20\,{\rm R_{\sun}}$, with a lower limit at $6\,{\rm R_{\sun}}$, and a virtually unconstrained upper limit. Summarizing we found that the flare occurs in a large-scale magnetic loop with half length $L\sim10-20\,{\rm R_{\sun}}=5-10\,R_{\star}$.

We note that there is no evidence of flare occultation by the disk or by the star itself because of stellar rotation, and that there is no evidence of higher absorption during the flare. This indicates that during its evolution the flaring loop always remains observable and that no significant portion of infalling material passes along the line of sight. 

\subsection{Dispersed spectra analysis and long term variability}
\label{highresspec}

The dispersed HEG and MEG spectra give access to individual line fluxes and allow us to derive detailed information on the X-ray emitting plasma such as the emission measure distribution ($EMD$), the element abundances, and the plasma density. The $EMD$ is a powerful tool to monitor the amount of plasma in small temperature bins and can only be derived from individual emission lines, not measurable in low resolution X-ray spectra. We measured line fluxes and performed our complete spectral analysis separately for the spectra gathered in each of the two observing segments. We can therefore explore, with the details offered by high resolution spectroscopy, variability on time scales of $\sim100$\,ks. Shorter time scales cannot be investigated with these diagnostic tools because of the limited signal to noise ratio. We also analyzed the spectrum collected during the entire observation so to obtain time-averaged plasma properties with the highest possible accuracy.

In order to increase the signal to noise ratio we added the MEG and HEG $1^{\rm st}$ order spectra, having first rebinned the HEG spectra to the MEG wavelength grid. With this procedure we obtained two dispersed spectra, one for each of the two observing segments, totaling 3750 and 4274 net counts, respectively. Figure.~\ref{fig:spec} shows the resulting spectra with labels identifying the strongest emission lines. The subsequent analysis was performed using the IDL package PINTofALE v2.0 \citep{KashyapDrake2000}, and the CHIANTI v4.2 atomic database \citep{YoungDelZanna2003}.

\subsubsection{Flux measurements}
\label{flux_measurements}

We measured the fluxes of the strongest emission lines in the observed X-ray spectra of V2129~Oph. Along with the line fluxes we also measured continuum fluxes in selected wavelength intervals free of significant line contamination (see below).

Line fluxes were obtained through direct integration of the binned spectra. For the integration we considered wavelength intervals centered around the predicted wavelength of each line\footnote{Given the discrete nature of the energy binning and the need to sum over an integer number of bins, the full width of the integration intervals is line-dependent, and ranges between 3 and 5 wavelength bins, corresponding to 2.2 and 3.4$\sigma$ of the line profiles.}. For each line we subtracted the continuum contribution, estimated from nearby wavelength intervals free of evident emission lines, and corrected the resulting flux for the fraction of the line-profile, assumed Gaussian, outside of the integration domain. Effective areas and exposure times were finally used to convert the observed line counts into fluxes. The observed net fluxes are listed, for each observing segment and for the entire observation, in Table~\ref{tab:fluxes}. For each line we indicate the ion generating it; line blends are indicated by listing the different ions contributing to the emission at that wavelength. We also list the temperature of maximum formation, $T_{\rm max}$, of each measured line (for line blends $T_{\rm max}$ is obtained from the sum of the emissivities of the blended lines). $T_{\rm max}$ ranges between 2 and 16\,MK. This range of maximum formation temperatures indicates that there is a significant amount of plasma at a few MK, that did not emerge from the analysis of the time averaged low resolution $0^{\rm th}$ order spectrum in Sect.~\ref{0ordermean}, of from the analysis of the $0^{\rm th}$ and heavily rebinned $1^{\rm st}$ order spectra in Sect.~\ref{shorttermvar}. In addition to the strong lines, we also included in Table~\ref{tab:fluxes} other well known lines, even if their fluxes are compatible with zero; upper limits are anyway useful constraints for the $EMD$ reconstruction.

\begin{figure}
\centering
\includegraphics[width=\columnwidth]{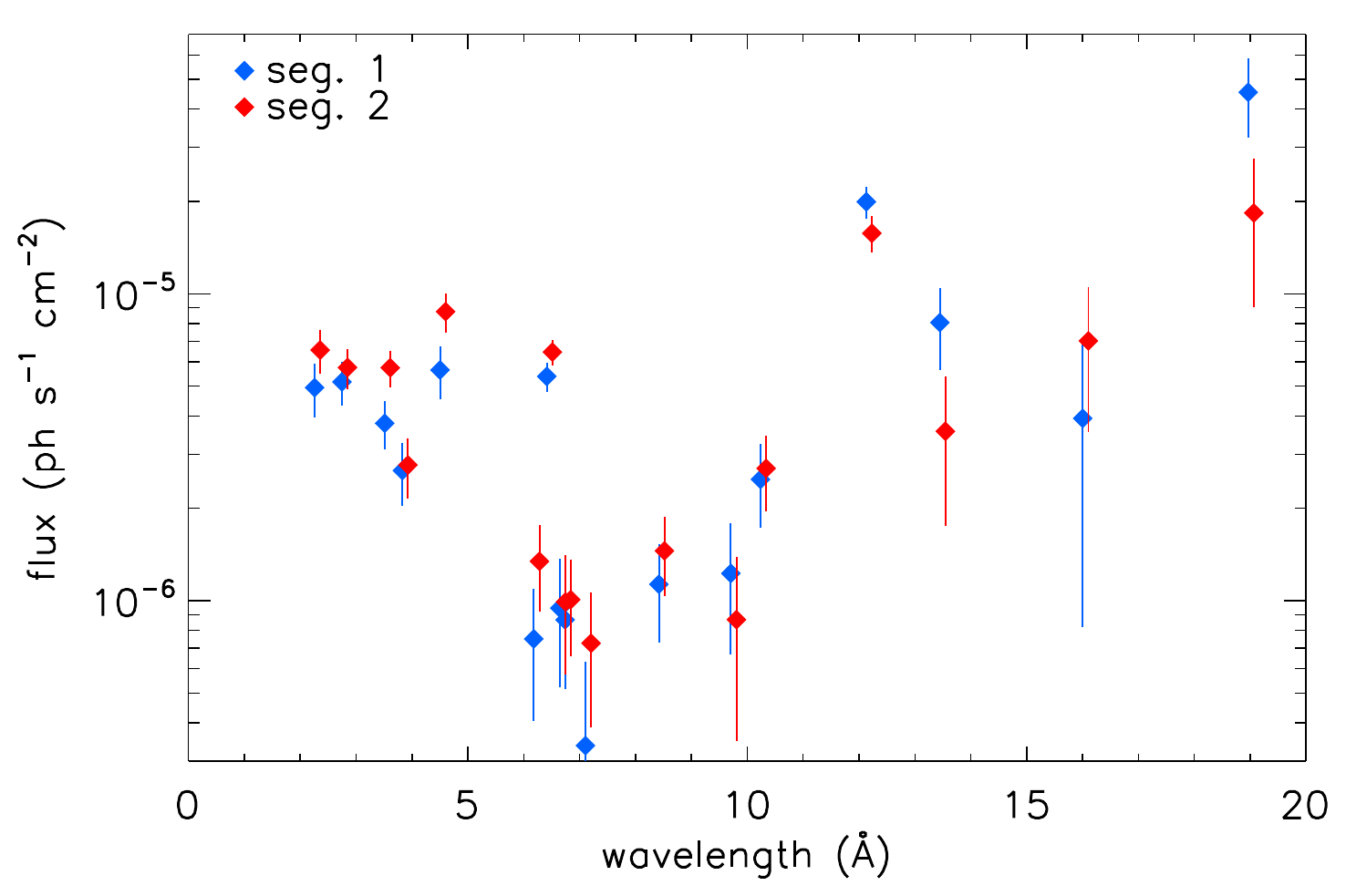}
\caption{Comparison of measured continuum and lines fluxes in the two observing segments. Only features detected at the $2\sigma$ level in at least one segment are plotted.}
\label{fig:compare_fluxes}
\end{figure}

In addition to line fluxes we also measured continuum fluxes in selected wavelength ranges. These measures are useful to constrain: a) the source metallicity\footnote{Line fluxes are proportional to abundances, while the continuum is not. The relative strength of line and continuum emission is therefore a measure of the plasma abundances.} and b) and the emission measure of the hottest plasma components\footnote{The major contribution to the continuum emission is due to bremsstrahlung, whose emissivity increases for increasing temperatures. The continuum emission level is therefore highly sensitive to the hottest plasma components.}. We measure the continuum emission of V2129~Oph in wavelength intervals that are free of significant line contributions (in particular we selected intervals in which the line contribution with respect to the continuum expected is below 10\%, assuming solar metallicities and for any temperature ranging between $10^{4}$ to $10^{8}$\,K). The continuum emission in these wavelength intervals was estimated by integrating the observed counts, and by subtracting the background contribution. The selected intervals and the corresponding measured fluxes are listed in Table~\ref{tab:fluxes}.

Figure~\ref{fig:compare_fluxes} shows a comparison of the measured fluxes (both lines and continuum) in the two observing segments. We only plot, as a function of wavelength, the fluxes of features detected at the $2\sigma$ level in at least one segment. Fluxes at short wavelengths tend to be, on average, higher in seg.~2 than in seg.~1, while the opposite results for fluxes at long wavelengths (these differences are also marginally observable in the plot of the two spectra, Fig.~\ref{fig:spec}). In many cases this difference is significant at the $1\sigma$ level. The analysis of the low resolution spectra has excluded significant variability of the absorption, hence we can reasonably ascribe differences in line fluxes to differences in the $EMD$.

The two coolest lines detected at the $2\sigma$ level in at least one segment are the \ion{O}{viii} Ly$\alpha$ line at 18.97\,\AA~and the \ion{Ne}{ix} resonance line at 13.45\,\AA, mostly emitted by plasma at $3-4$\,MK. These lines show flux differences between the two segments of 1.7 and 1.5$\sigma$, with seg. 1 fluxes higher than seg. 2. Combining the two differences we obtain a very low probability (0.3\%) that the $EM$ of the plasma at $3-4$\,MK producing this lines has not changed. Continuum fluxes, that probe variations of the hottest plasma components, are systematically higher during seg.~2 than during seg.~1, firmly indicating that the $EM$ at $T>10$\,MK is higher during seg. 2.

\subsubsection{Emission Measure Distribution}
\label{emd}

\begin{figure}
\centering
\includegraphics[width=\columnwidth]{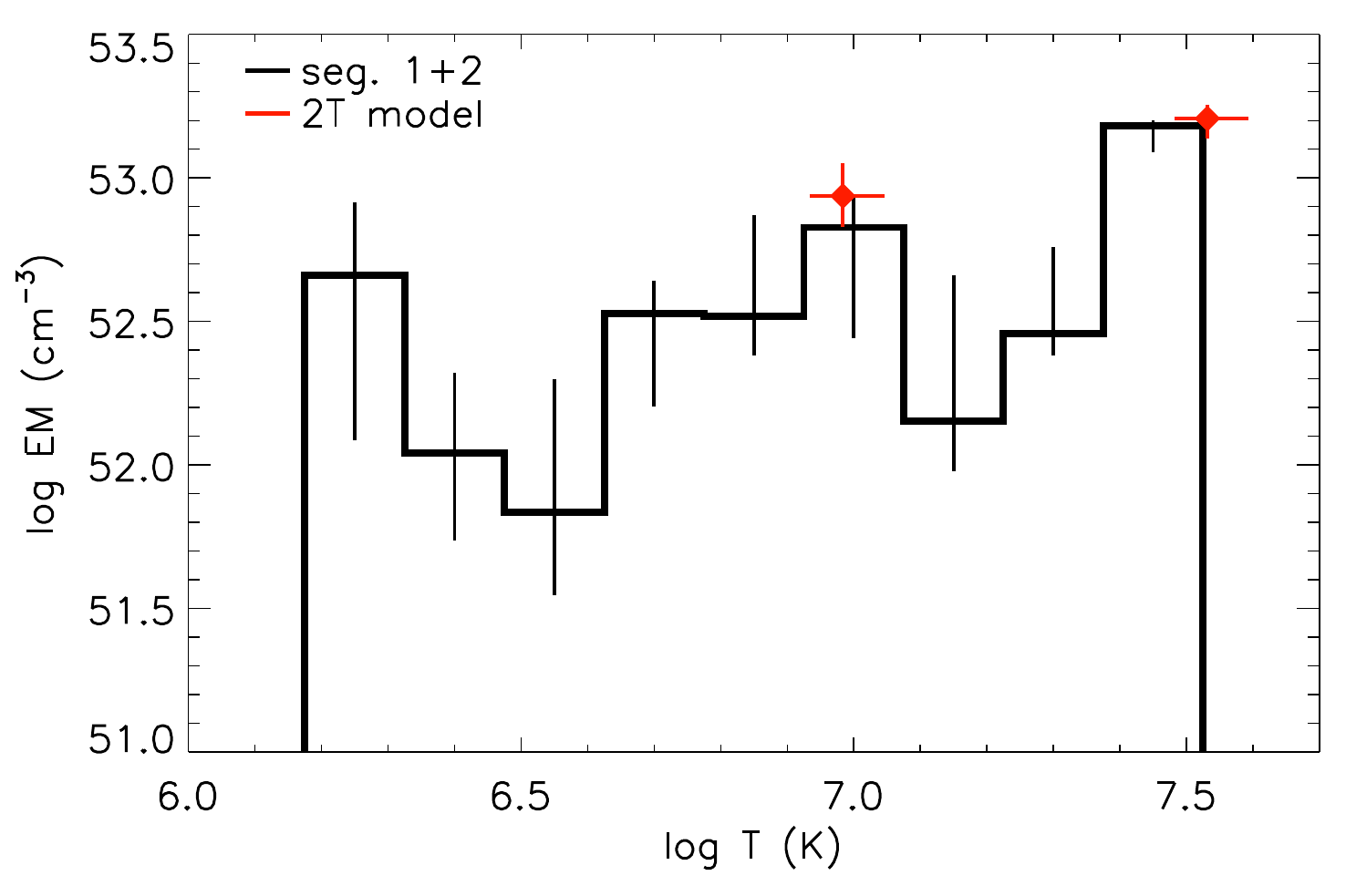}
\caption{$EMD$ obtained from the flux measurements of the whole observation (seg.~1+2).}
\label{fig:emd12}
\end{figure}

\begin{figure}
\centering
\includegraphics[width=\columnwidth]{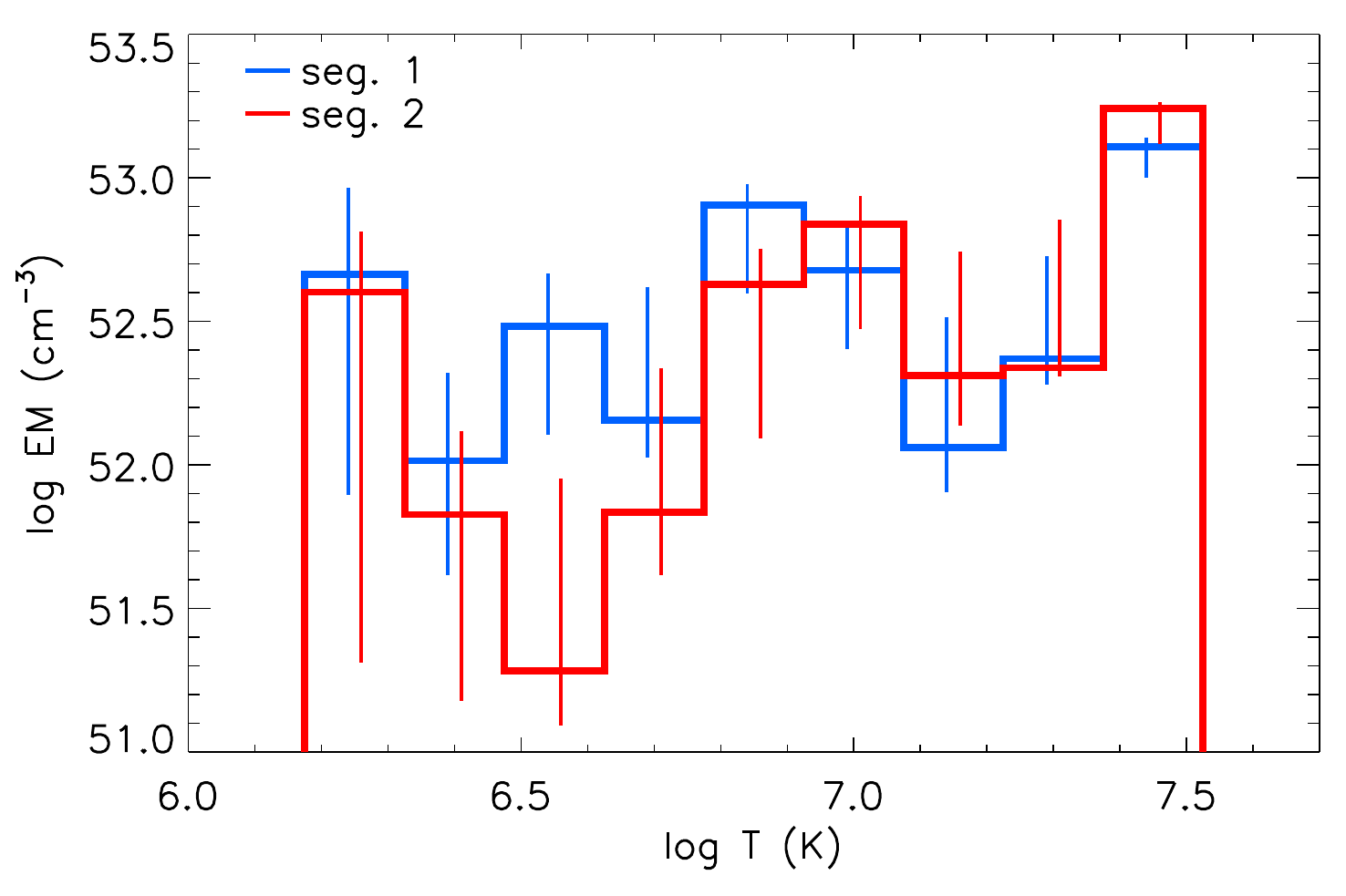}
\caption{$EMD$ obtained from the flux measurements of the two observing segments.}
\label{fig:emd1vs2}
\end{figure}

We derived the $EMD$ and abundances of the plasma responsible for the X-ray emission of V2129~Oph by applying the Markov-Chain Monte Carlo method of \citet{KashyapDrake1998} to the measured continuum and line fluxes in Table~\ref{tab:fluxes}. Measured fluxes were converted into unabsorbed fluxes adopting the hydrogen column density $N_{\rm H}$ of $8\times10^{20}\,{\rm cm^{-2}}$ derived from the $0^{\rm th}$ order spectrum (see Sect.~\ref{0ordermean}). For the $EMD$ and abundance reconstruction we considered all the measured fluxes discarding only the lines that depend on plasma density, i.e. the intercombination and forbidden lines of He-like triplets.

Starting from the flux measurements and considering the peak temperature at which each line is produced (see Table~\ref{tab:fluxes}) we reconstructed the $EMD$ over a regular logarithmic temperature grid with $0.15$\,dex bins in the $\log T$ range between $6.2$ and $7.5$. Abundances were estimated for those elements for which at least one line was measured. Note that the reconstructed $EM$ values associated to bins at the boundary of the chosen temperature domain may be overestimated: artificially limiting the temperature range of the $EMD$ reconstruction forces to zero the $EMD$ outside this range. Plasma at higher and lower temperatures may however contribute to some of the measured fluxes, and thus causing the $EM$ values at the boundary bins to be overestimated in order to account for the observed fluxes.

The $EMD$ and abundances derived from the whole observation are reported in Table~\ref{tab:model} and the $EMD$ is also plotted in Fig.~\ref{fig:emd12}. Figure~\ref{fig:emd12} also shows the 2 temperature model obtained from the $0^{\rm th}$ order spectrum of the whole observation (see Sect.~\ref{0ordermean}). We note that the two components of this 2-$T$ model agree very well with the two hottest peaks of the $EMD$.

\begin{table}
\renewcommand{\baselinestretch}{1.3}
\caption{EMD and abundances of V2129~Oph.}
\label{tab:model}
\normalsize
\begin{center}
\begin{tabular}{cccc}
\hline\hline
 & seg.~1 & seg.~2 & seg.~1+2 \\
\hline
$\log T$~(K) & $\log EM$ & $\log EM$ & $\log EM$ \\
\hline
 $  6.25$ & $ 52.67^{+  0.30}_{-  0.77}$ & $ 52.60^{+  0.21}_{-  1.29}$ & $ 52.66^{+  0.25}_{-  0.57}$ \\
 $  6.40$ & $ 52.01^{+  0.31}_{-  0.40}$ & $ 51.83^{+  0.29}_{-  0.65}$ & $ 52.04^{+  0.28}_{-  0.30}$ \\
 $  6.55$ & $ 52.48^{+  0.18}_{-  0.38}$ & $ 51.28^{+  0.67}_{-  0.19}$ & $ 51.84^{+  0.46}_{-  0.29}$ \\
 $  6.70$ & $ 52.16^{+  0.46}_{-  0.13}$ & $ 51.83^{+  0.50}_{-  0.22}$ & $ 52.53^{+  0.12}_{-  0.32}$ \\
 $  6.85$ & $ 52.91^{+  0.07}_{-  0.31}$ & $ 52.63^{+  0.13}_{-  0.54}$ & $ 52.52^{+  0.35}_{-  0.14}$ \\
 $  7.00$ & $ 52.68^{+  0.15}_{-  0.27}$ & $ 52.84^{+  0.10}_{-  0.37}$ & $ 52.83^{+  0.10}_{-  0.39}$ \\
 $  7.15$ & $ 52.06^{+  0.46}_{-  0.16}$ & $ 52.31^{+  0.43}_{-  0.18}$ & $ 52.15^{+  0.51}_{-  0.17}$ \\
 $  7.30$ & $ 52.37^{+  0.36}_{-  0.09}$ & $ 52.34^{+  0.52}_{-  0.03}$ & $ 52.46^{+  0.30}_{-  0.08}$ \\
 $  7.45$ & $ 53.11^{+  0.03}_{-  0.11}$ & $ 53.24^{+  0.02}_{-  0.12}$ & $ 53.18^{+  0.02}_{-  0.09}$ \\
\hline
\multicolumn{4}{c}{Abundances} \\
\hline
\multicolumn{4}{c}{  O$\,=  0.61^{+  0.21}_{-  0.21}$ \hfill Ne$\,=  1.09^{+  0.29}_{-  0.26}$ \hfill Mg$\,=  0.15^{+  0.04}_{-  0.04}$} \\
\multicolumn{4}{c}{ Si$\,=  0.15^{+  0.03}_{-  0.04}$ \hfill  S$\,=  0.38^{+  0.14}_{-  0.32}$ \hfill Fe$\,=  0.13^{+  0.04}_{-  0.04}$} \\
\hline
\end{tabular}
\end{center}
\renewcommand{\baselinestretch}{1.0}
\small
Abundances are in solar units \citep{AsplundGrevesse2005}. EMD values are in ${\rm cm^{-3}}$ and have been derived assuming that V2129~Oph is located at 120\,pc.
\normalsize
\end{table}
\normalsize

We then derived $EMD$ and abundances of the emitting plasma considering separately the two observing segments. For both segments unabsorbed fluxes were evaluated assuming $N_{\rm H}=8\times10^{20}\,{\rm cm^{-2}}$, because no evidence of absorption variability was found (see Sect.~\ref{timeresspec}). No significant variation in plasma abundances can be discerned  between seg.~1 and 2, indicating that the observed variations of line fluxes can be ascribed to $EMD$ variations only. We therefore repeated the $EMD$ reconstruction for seg.~1 and 2 freezing the plasma abundances to the values of the seg.~1+2 model, and leaving as free parameters the $EM$ values. The thus derived $EMD$ for the two segments are reported in Table~\ref{tab:model}, and plotted in Fig.~\ref{fig:emd1vs2}. As a consequence of the observed systematic differences in measured line fluxes, a  difference is seen in the cool plasma component\footnote{Note that the corresponding models obtained without fixed abundances, have $EMD$ values almost identical to those obtained by fixing them, but the associated $EMD$ uncertainties are larger and the differences at low temperatures are not significant.}, with $EMD_1>EMD_2$ at the 1.5$\sigma$ level in the bin centered on $\log T = 6.55$ ($T\sim3-4$\,MK). Note that the significance of the time variability is lower for the $EMD$ than for the line fluxes (see Sect.~\ref{flux_measurements}) because the $EMD$ reconstruction procedure involves further uncertainty sources \citep{KashyapDrake1998}. However the high confidence level inferred for the variation of the flux of cool lines indicates that the $EMD$ variability is indeed real.

\subsubsection{Electron density}

The plasma density $N_{\rm e}$ can be constrained, from the high resolution X-ray spectroscopy, by analyzing the He-like line triplets, and, in particular, the flux ratio between the intercombination line, $i$, and the forbidden line, $f$ \citep{GabrielJordan1969}. Increasing plasma densities result in decreasing $f/i$ ratios. The intensities of the $i$ and $f$ lines depend on plasma density because these lines originate from radiative decay of metastable states (long lifetime states), whose depopulation is favored by increasing collision rate (i.e. increasing density), while the resonance lines originate from excited states depopulated only by photon emission processes.

Among the He-like triplets in the wavelength range covered by the {\it Chandra}/HETGS instrument, we identified and measured the He-like line triplets of \ion{Si}{xiii} (at 6.65, 6.69, 6.74\,\AA), \ion{Mg}{xi} (at 9.17, 9.23, 9.31\,\AA), and \ion{Ne}{ix} (at 13.45, 13.55, 13.70\,\AA). The analysis of these triplets can constrain the electron density of plasma at the formation temperatures of these ions, $\sim10$, $\sim6$, and $\sim4$\,MK, respectively. Measured line fluxes for these triplets are reported in Table~\ref{tab:fluxes}.

The \ion{Si}{xiii} triplet has, in all of our spectra, $i$ line fluxes compatible with zero within 1$\sigma$, providing therefore only a lower limit for the $f/i$ ratio. We therefore infer that plasma at $T\sim10\,MK$ has a density lower than $\sim10^{12}\,{\rm cm^{-3}}$, during both seg.~1 and seg.~2. The \ion{Mg}{xi} triplet, mostly emitted by plasma at 6\,MK, has both $i$ and $f$ line fluxes compatible with zero in all the spectra, thus providing no constraint on plasma density.

\begin{figure}
\centering
\includegraphics[width=\columnwidth]{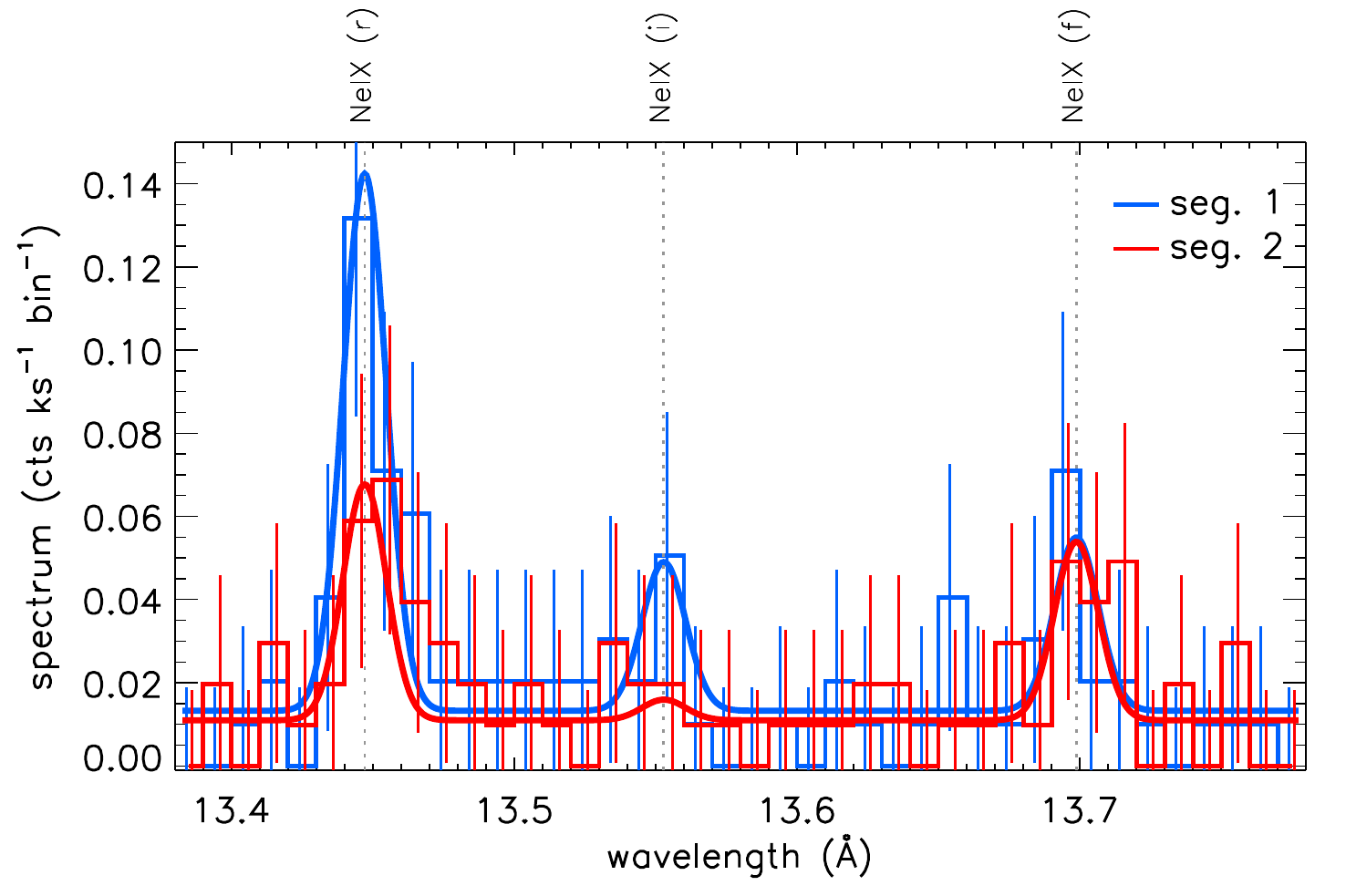}
\caption{Spectra of the two observing segments in the \ion{Ne}{ix} triplet region. Observed spectra are rebinned by a factor 2 (bin size~$=0.01$\,\AA). Model spectra are obtained using line and continuum measurements (see Sect.~\ref{flux_measurements}) and assuming a gaussian line spread function.}
\label{fig:NeIXtrip}
\end{figure}

\begin{figure}
\centering
\includegraphics[width=\columnwidth]{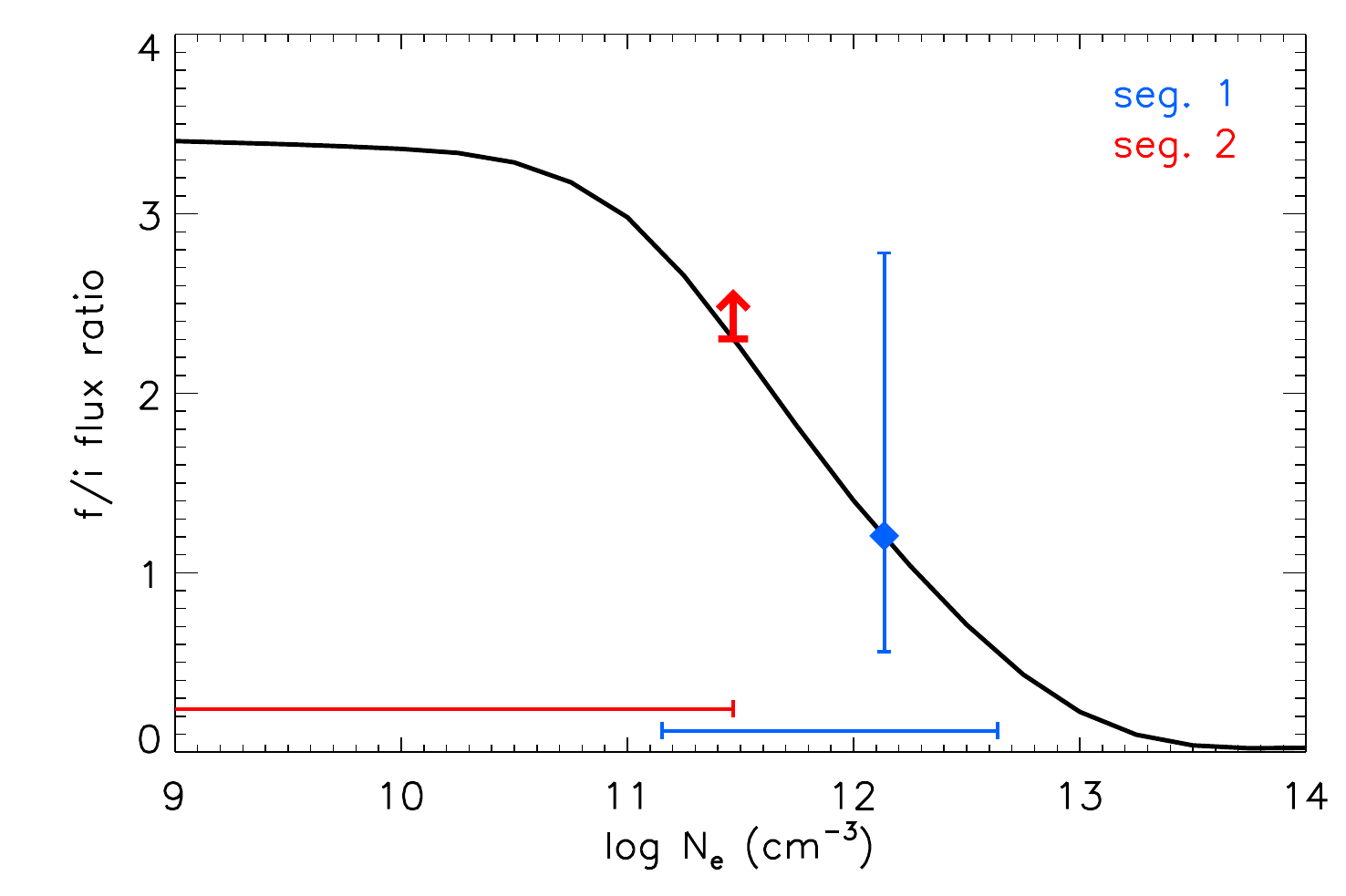}
\caption{Forbidden $f$ to intercombination $i$ line ratio of the \ion{Ne}{ix} triplet vs electron density. Black curve represents the $f/i$ predicted ratio. Observed $f/i$ values of segments 1 and 2, and the corresponding density ranges, are indicated in blue and red respectively.}
\label{fig:NeIXfiratio}
\end{figure}

In Fig.~\ref{fig:NeIXtrip} we show the lines of the \ion{Ne}{ix} triplet corresponding to the two observing segments. These lines are produced by plasma at $\sim4$\,MK. In the plot we also show noise-free spectra reconstructed on the basis of the measured continuum and line counts. The overall flux of the triplet is clearly lower in the second observing segment than in first  (see also Table~\ref{tab:fluxes}). 

To obtain an accurate estimation of the $f/i$ ratio uncertainties we considered the measured counts of $i$ and $f$ lines, as well as those used to determine the continuum level. Assuming that these quantities follow Poisson distributions we derived through Monte Carlo simulations the distribution of the $f/i$ flux ratio, and hence inferred the $f/i$ mean value and the quantile corresponding to $\pm1\sigma$. With this procedure, and having also corrected for interstellar absorption, we obtained $f/i=1.2^{+1.6}_{-0.6}$ and $f/i>2.3$ for seg.~1 and 2, respectively. The comparison of the observed $f/i$ values with the predicted $f/i$-ratio vs. plasma density function\footnote{The predicted ratios, other than on density, also depend, very weakly, on the plasma $EMD$. In deriving the $f/i$-ratio vs. density relation we adopted the mean $EMD$ for whole observation. Using the individual $EMD$ derived for the two segments, however, produces negligible differences in the inferred $\log N_{\rm e}$, smaller than $0.01$\,dex.} is shown in Fig.~\ref{fig:NeIXfiratio}. From this relation we derive: $\log N_{\rm e} = 12.1^{+0.5}_{-1.0}$ and $\log N_{\rm e} < 11.5$ for seg.~1 and 2, respectively, with $N_{\rm e}$ in ${\rm cm^{-3}}$.

\section{Main results}
\label{results}

The {\it Chandra} observation of V2129~Oph has revealed that it is an X-ray bright source. The average X-ray luminosity, derived from our mean $EMD$, is $L_{\rm X}=2.7\times10^{30}\,{\rm erg\,s^{-1}}$ in the $0.5-10.0$\,keV band. This $L_{\rm X}$, combined with the bolometric luminosity of V2129~Oph \citep[$\log L_{\rm bol}/L_{\sun}=0.15$,][]{DonatiBouvier2011}, yields $\log L_{\rm X}/L_{\rm bol} = -3.3$, a value that puts V2129~Oph among the X-ray brightest accreting T~Tauri stars when compared to CTTS in the ONC \citep{PreibischKim2005}. The X-ray emission from V2129~Oph is produced by significant amounts of plasma spanning a broad temperature range, at least from 2\,MK up to  $\sim30$\,MK, as evidenced by the reconstructed $EMD$ (see Fig.~\ref{fig:emd12}). In particular the plasma at a few MK, monitored during the first half of the {\it Chandra} observation, shows high density, $\log N_{\rm e} \approx 12.1$, consistent with results for almost all the other CTTS with measured densities, supporting the idea that the presence of high density cool plasma is directly linked to the accretion phenomenon. 

\subsection{X-ray variability}

The X-ray emission of V2129~Oph is variable on short time scales (hours). In the 200\,ks {\it Chandra} observation we detected a clear flare, $\sim$10\,h long, causing an increase in the X-ray luminosity by a factor $\sim3$. This flare occurred in a long loop, with an analysis of the rise phase giving a best value of the loop half length of $\sim5-10\,R_{\star}$, with a lower limit of $3\,R_{\star}$, and an unconstrained upper limit.

Beside this clear flare, all the observation is characterized by variability similar to that typically produced by coronal activity, i.e. with clearly correlated X-ray emission level and spectral hardness. This short term variability is in fact mainly produced by the hottest ($T\approx34$\,MK) plasma components.

The analysis of the high resolution spectra collected during the two segments allowed us to search for variability of the X-ray emission observed at two different phases during stellar rotation and, therefore, emerging from either different magnetospheric structures (corona and accretion shock) or from the same structures seen with different viewing geometries. We stress that the analysis of the high resolution spectra allows us to probe the coolest plasma components. This comparison yielded two main results.

\begin{enumerate}

\item {\it Cool plasma $EM$ variability}. Lines at long wavelengths were significantly enhanced during seg.~1 with respect to seg.~2. This finding is apparently not related to variation in the absorbing column since no such variability is derived from the XSPEC analysis of the heavily rebinned $0^{\rm th}+1^{\rm th}$ order spectra. The $EMD$ of the two observing segments indeed indicates that this difference originates from a difference in the amount of observable emission measure at low temperatures. In particular we observe a difference in the $EM$ at $T\approx3-4$\,MK ($\log T = 6.55$), with $EMD_{1}$ being higher than $EMD_{2}$, at the 1.5$\sigma$ level, by more than a factor of 10. The difference in the amount of emission measure at a few MK is likely not related to flaring activity, since, as indicated by the short time-scale analysis, flares contribute to the $EMD$ at much hotter temperatures than 4\,MK (see Fig.~\ref{fig:2Tem_vs_time}).

\item {\it Plasma density variability}. During the two observing segments we found different constraints on the density of the plasma component at a few MK. In segment 1 the cool plasma density can be constrained between $1.3\times10^{11}$ and $4\times10^{12}\,{\rm cm^{-3}}$, while during seg.~2 the density diagnostic is compatible with a low density plasma, $N_{\rm e}<3\times10^{11}\,{\rm cm^{-3}}$. Although the statistical significance of this difference is low, V2129~Oph, is the first CTTS where apparent variations of the plasma density over time scales of $\sim100$\,ks have been observed.

\end{enumerate}

\section{Discussion}
\label{disc}

We now discuss the implications of our observational results on the X-ray emission of V2129~Oph, taking advantage of the constrains on the properties and geometry of accretion shock region and on the stellar magnetic field, obtained by us simultaneously with the X-ray observation \citep{DonatiBouvier2011}.

The analysis of X-ray data allows us to derive the main properties of the coronal plasma and of the plasma heated in accretion shocks, as discussed below. These results will also provide constraints for future works dedicated to magnetic field extrapolation, aimed at modeling the coronal plasma and the accretion stream structures.

\subsection{Coronal emission}

The broad band X-ray luminosity of V2129~Oph is dominated by emission from hot plasma components, most likely of coronal origin. In 1991 the {\it ROSAT} satellite detected V2129~Oph \citep{CasanovaMontmerle1995} with an X-ray luminosity of $2.5\times10^{30}\,{\rm erg\,s^{-1}}$, almost identical to that measured with {\it Chandra} in 2009. V2129~Oph thus appears to have a stable X-ray emission, hence coronal activity level, over time scales of $\sim20$\,yr.

V2129~Oph shows X-ray variability over short time scales (Fig.\,\ref{fig:lc}). This variability, being mainly due very hot plasmas, can be attributed to coronal activity. The X-ray light curve of the first observing segment is characterized by an obvious flare. We constrained the half length of the flaring loop, deriving $5-10\,R_{\star}$ as best value (with a lower limit of $3\,R_{\star}$ and with an unconstrained upper limit). Considering its length and the inner disk radius ($\sim7.2\,R_{\star}$), we cannot exclude the possibility that this loop has one footpoint anchored to the inner circumstellar disk. Such flaring coronal structures connecting star and inner disk were already suggested in young stars of the Orion Nebula Cluster \citep{FavataFlaccomio2005,GetmanFeigelson2008}, although we note that the presence of a disk is not a prerequisite for such large flaring loops to exist. They are also found to occur on stars which show no evidence for disks \citep{GetmanFeigelson2008,AarnioStassun2010}.

The second segment of the {\it Chandra} observation shows a trend in the first part of the segment that might be interpreted as the decay phase of a flare, followed in the second part by an increase of the count rate and of the hardness ratio that might be associated the rise phase of another flare.

Alternatively the variability might be due to modulation of some emission feature on the stellar surface as the star rotates. We note that the observed dip in the light curve coincides with the passage along the line of sight of the large region of strong positive radial magnetic field (in red in Fig~\ref{fig:lc}), which is also coincident with an extended dark photospheric spot \citep[not shown, cf. ][]{DonatiBouvier2011}. Although we can only speculate that the dip in the X-ray light curve is associated with the passage of this region, it is not unconceivable that the X-ray emitting corona, assumed to be confined by relatively small-scale closed magnetic loops, is disrupted in this region of strong magnetic field likely connecting the star with the inner disk (as it is certainly true for the field in the accretion spot that lies within the region). Figure~\ref{fig:2Tem_vs_time} might further indicate that while this region removes a significant fraction of visible {\it hot} (34\,MK) plasma, it leaves the {\it cool} (9.6\,MK) plasma relatively unaffected thus indicating that the two are not co-spatial.

\subsection{High density cool plasma and accretion shock(s)}

We found that, during the first segment of our {\it Chandra} observation, the plasma at a few MK has a high density ($\log N_{\rm e} = 12.1$). Therefore V2129~Oph is a further case confirming that in CTTS cool plasma has high density. This plasma component is intrinsically different from the coronal plasma observed in non-accreting active stars. The commonly accepted model suggests that this plasma component is material heated in the accretion shock. In fact, the high infall velocity of the accreting material naturally justifies material at such temperatures, of a few MK, and densities.

Plasma heated in accretion shocks is located at the base of the accretion stream, just behind the shock surface. Models predict that this accreted material, passing through the shock surface, strongly reduces its infall velocity, but continues its downward motion across the stellar atmosphere, meanwhile cooling down and
radiating X-rays \citep{GuntherSchmitt2007,SaccoArgiroffi2008,OrlandoSacco2010}.

The association of the cool plasma component at $\sim3-4$\,MK of V2129~Oph with the accretion shock is greatly reinforced if we consider that this temperature is almost exactly what we actually expect for the post-shock plasma. The measured strength of the magnetic field of V2129~Oph implies that inner disk is truncated at $7.2\,R_{\star}$ \citep{DonatiBouvier2011}. A free fall from that distance implies a pre-shock velocity of $460\,{\rm km\,s^{-1}}$, and, as a consequence, a post-shock temperature of $3$\,MK.

\begin{figure*}
\centering
\includegraphics[width=18cm]{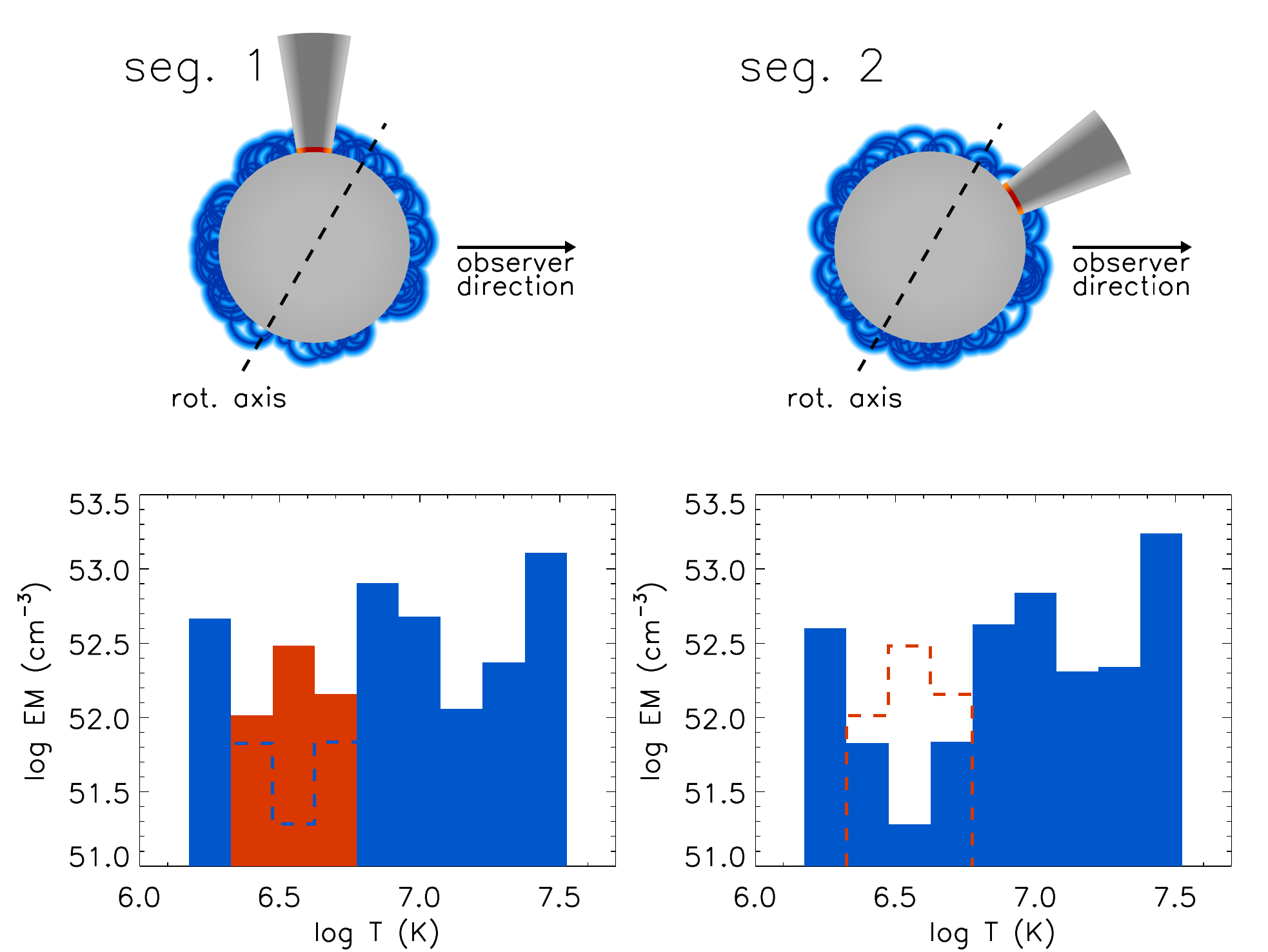}
\caption{{\it Upper panels}: Cartoon describing the spatial distribution of the two X-ray emitting plasma components of V2129~Oph. Left and right cartoons correspond to the two viewing angles of the two {\it Chandra} observing segments, with observer in the rightward direction. Red region marks the post-shock high density plasma at $3-4$\,MK, blue regions indicate low density coronal plasma with $T$ ranging from 2 up to $\sim30$\,MK. {\it Lower panels}: $EMD$ corresponding to the two observing segments (see Sect.~\ref{emd}). Red $EM$ bins symbolize observed $EM$ values ascribed to post-shock plasma, while blue bins those accounting for coronal plasma. During seg.~1 the pre-shock material does not block the view of both post-shock and coronal plasma: the $EMD$ is therefore the superposition of the $EMD$ two components. During seg.~2 the pre-shock material almost completely absorbs the X-rays of the post-shock plasma emitted toward the observer, while coronal emission is mostly unaffected: in this case all the X-rays detected are those produced by coronal plasma; the reconstructed $EMD$, being only those of coronal plasma, hence misses the high $EM$ values at $3-4$\,MK (reported for comparison with a dashed red line in the $EMD$ plot).}
\label{fig:cart}
\end{figure*}

We will therefore assume that X-rays from V2129~Oph are emitted by two plasma components: a) the high density plasma component at $3-4$\,MK, heated in the accretion shock and hence located at the base of the accretion stream, and b) a low density corona, with $T$ ranging from 2 up to a $\sim30$\,MK. We will try to interpret the time variability of the $EM$ at $\sim3-4$\,MK and, possibly, of the measured plasma density that we observed between our two observing segments in terms of different viewing geometry of the accretion shock and funnel due to stellar rotation. Figure~\ref{fig:cart} displays a cartoon describing the system and the viewing geometry during the two observing segments.

From a theoretical point of view, it has been debated whether X-rays emitted by shock-heated plasma in CTTS are able to escape from the shock region, and hence be detected \citep{Drake2005,SaccoOrlando2010}. Their escape probability depends on the height of the shock surface on the stellar atmosphere and on the direction of emission. In fact, X-rays emitted in the direction of the accretion stream could be absorbed by the pre-shock infalling material, depending on the accretion stream density and dimensions along the line of sight; X-rays emitted perpendicularly to the accretion funnel could be absorbed by the surrounding stellar atmosphere, depending on the sinking of the hot post shock in the stellar atmosphere \citep{Drake2005,SaccoOrlando2010}. The amount of X-rays produced by the accreton shock able to escape in the different directions may be estimated through detailed 3D modeling of the shock region, which is beyond the scope of this paper.

In the case of V2129~Oph, we can assume that the post-shock plasma is co-spatial with the the optical hot spot, whose location we have determined through Zeeman-Doppler imaging. As shown in Fig.~\ref{fig:lc} and outlined in the cartoon of Fig.~\ref{fig:cart}, this region remains visible during both {\it Chandra} observing segments, but with distinctly different viewing geometries: near the stellar limb during seg.~1 and nearer the center of the stellar disk during seg.~2.

During seg.~1 the accretion funnel is observed sideways. In this configuration, X-rays produced by the shock heated plasma can easily be observed because they can escape from the lateral part of the funnel. In fact the accretion shock characteristics of V2129~Oph, i.e. a pre-shock velocity of $460\,{\rm km\,s^{-1}}$, a low metallicity, and a pre-shock density of $\sim3\times10^{11}\,{\rm cm^{-3}}$ (since the accretion shock is expected to be a strong shock, the density of the pre-shock material should be one fourth of the measured post-shock density), imply that the hot post-shock region is high enough on the stellar surface to avoid significant absorption from the surrounding chromosphere \citep{SaccoOrlando2010}. Of course X-rays from coronal plasma are also observable. The emerging X-ray spectrum and the resulting $EMD$ are therefore the superposition of the two contributions, with the post-shock component dominating the $EMD$ at $3-4$\,MK, and the corona dominating at higher temperatures. The observed X-ray spectrum of seg.~1 is in fact compatible with an overall low absorption, and with high density for the plasma at $\sim4$\,MK.

Conversely, during seg.~2 the hot spot almost faces the observer. Pre-shock infalling material is thus located between us and the post-shock region at the base of the accretion funnel (see Fig.~\ref{fig:cart}). The X-rays emitted in our direction by post-shock plasma thus enter the pre-shock accretion stream. Even considering that our line of sight intersects a small fraction of the accretion funnel, i.e. 15\% of stellar radius, the resulting hydrogen column density is $\sim7\times10^{21}\,{\rm cm^{-2}}$ (having assumed a pre-shock density of $3\times10^{11}\,{\rm cm^{-3}}$). Such a large column density would absorb 90\% of X-rays at $\sim13.5$\,\AA~(the wavelength of the \ion{Ne}{ix} triplet). In this viewing geometry essentially no X-ray emitted by the shock heated plasma can be observed. Therefore during seg.~2 only coronal X-rays can be observed, because they are not likely to be blocked by the pre-shock material. The emerging X-ray spectrum is therefore almost completely due to coronal plasma. The reconstructed $EMD$ of seg.~2, hence misses the high $EM$ values at $3-4$\,MK, and the the seg.~2 \ion{Ne}{ix} triplet reveals low density plasma, compatible with a coronal origin. In this geometry the emerging X-ray spectrum suffers again the same low absorption of seg.~1.

We notice that the lowest temperature bin of the $EMD$ (at 2\,MK), even if poorly constrained, appears constant between the two segments. This does not contradict the above picture for two reasons: 1) the error bars on the $EM$ might hide even a large variation, 2) HD simulations predict that accretion shocks onto CTTS produce sharply peaked $EMD$ \citep{SaccoOrlando2010} that might not contribute significantly to this $EM$ bin.

The proposed scenario is able to explain the observed variations of the $EMD$ and density with significant variations of the extinction suffered by the X-rays from the accretion shock. Note that this is compatible with the lack of variability in the absorption measured from the X-ray data. In fact an increase of the absorption in the X-rays of the post-shock plasma only cannot be detected because of the presence of the underlying weakly absorbed X-ray coronal emission.

Note that in the optical band, the accretion indicators (e.g. the \ion{Ca}{ii} and \ion{He}{i} lines) reach their maximum at phases $0.55-0.7$, corresponding to rotational phases at which the hot spot is directed toward us. We have here observed that the opposite happens for the accretion-driven X-ray emission: it is maximum when the hot spot is near the stellar limb and minimum when the hot-spot is facing us. We can make sense of these two opposite behaviors because the infalling material should be almost completely in the gas phase. Gas is indeed transparent to optical radiation, so that the \ion{Ca}{ii} and \ion{He}{i} emission can reach us through the funnel, but can easily absorb soft X-rays with energies $E<1$\,keV, such as those produced in the accretion shock.

We have argued that the plasma at $3-4$\,MK observed on V2129~Oph is likely to be shock-heated material. The high density, the time variability of its $EM$ in agreement with this picture, and the coincidence of the temperature with the expectation for the post-shock material reinforce this interpretation. In this likely hypothesis we can infer the properties of the shock heated plasma, and in particular the accretion rate, needed to explain the observed soft X-ray flux. From the observed flux of the \ion{Ne}{ix} resonance line measured during seg.~1, and assuming a pre-shock velocity of $460\,{\rm km\,s^{-1}}$ (compatible with a free fall from the estimated inner disk position), we derive an accretion rate of $7\times10^{-11}\,{\rm M_{\sun}\,yr^{-1}}$. This is lower by a factor 10 with respect to the accretion rate derived from optical lines, $6\times10^{-10}\,{\rm M_{\sun}\,yr^{-1}}$ \citep{DonatiBouvier2011}. This underestimation of $\dot{M}$ derived from X-rays is common for CTTS observed with high resolution X-ray spectroscopy \citep{ArgiroffiMaggio2009,CurranArgiroffi2011}. As suggested by \citet{GuntherSchmitt2007} and investigated by \citep{SaccoOrlando2010}, this discrepancy could be explained assuming that the accretion stream is not uniform in density and/or in velocity, and that therefore the portion of the accretion stream producing observable X-rays is a small fraction of the whole stream. Moreover \citet{CurranArgiroffi2011}, comparing mass accretion rates of CTTS derived from optical diagnostics with those derived from high resolution X-ray spectra, argued that, even if X-rays provide significantly underestimated values, a weak correlation between the two quantities may exist. The values that we obtain for V2129~Oph support this tentative correlation.

Finally, having argued that, at least during seg.~1, the $EMD$ bins at low temperature are due to the accretion-shock and those at high temperature to coronal plasma, we estimate separately the X-ray luminosities of the two components. We do so by assuming that the $EMD$ below and above $\log T=6.8$ are entirely due, respectively, to shock heated material and to coronal plasma. From the $EMD$ of seg.~1, the resulting $L_{\rm X}$ are $3.4\times10^{29}$ and $2.2\times10^{30}\,{\rm erg\,s^{-1}}$, for the shock heated and the coronal plasma, respectively. Even when we do observe X-rays from shock heated material (seg.~1), the coronal X-ray luminosity is thus $\sim90$\% of the total $L_{\rm X}$.

A similar conclusion can be reached by comparing the $EMD$ reconstructed from individual line and continuum fluxes, with the 2 temperature model obtained from the spectral fitting of the CCD-resolution ACIS spectrum (see Fig.~\ref{fig:emd12}). The temperatures in this latter model correspond to the two hottest peaks of the $EMD$ distribution, both ascribable to coronal plasma. This is a further confirmation that X-ray spectra of CTTS in the $0.5-10$\,keV band are dominated by coronal emission. The shock heated material is probed only by the softest part of the spectrum, or by emission lines produced by material at temperatures of few MK, that contribute little flux to the overall low-resolution spectra.

\begin{figure}
\centering
\includegraphics[width=\columnwidth]{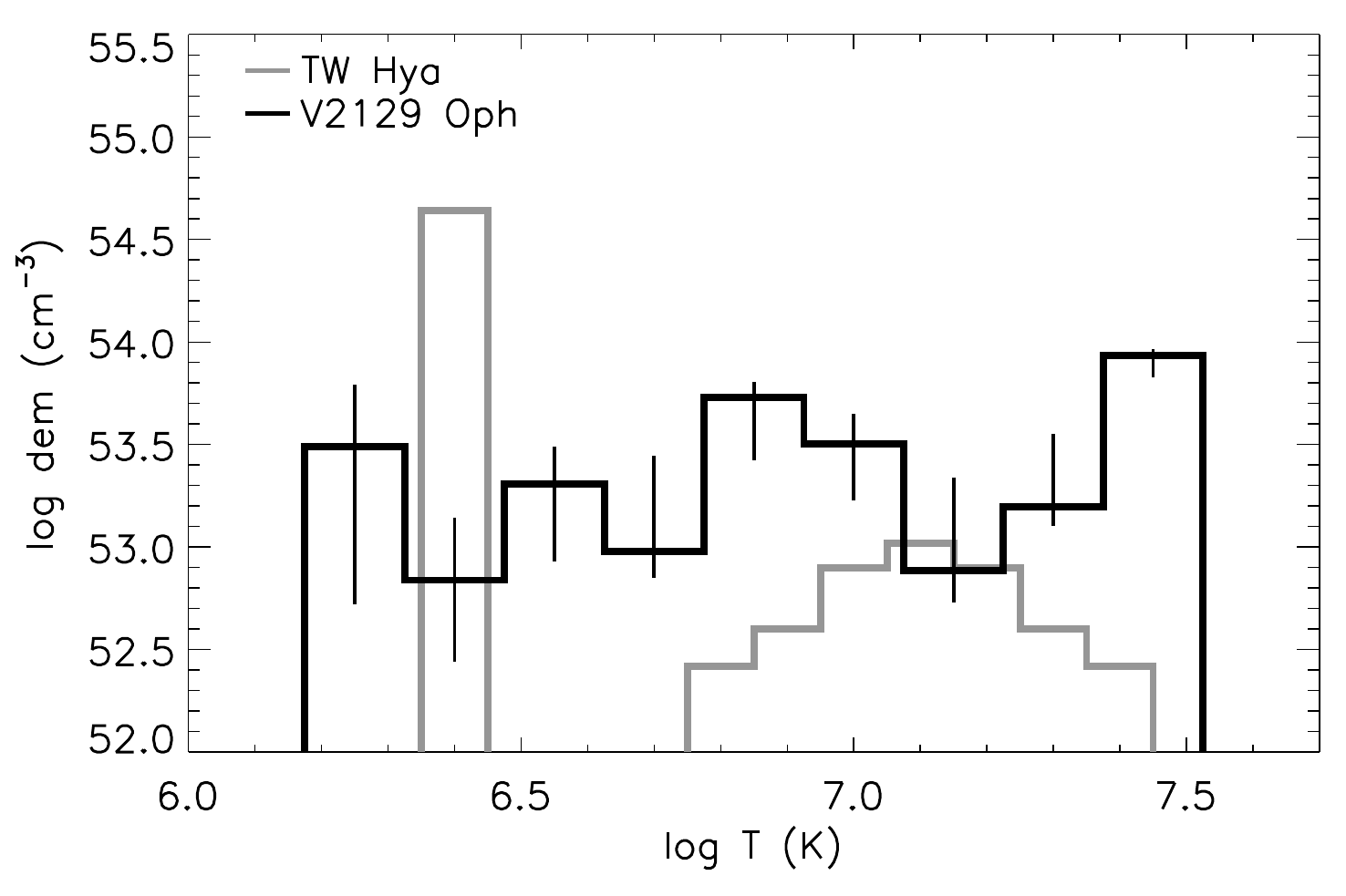}
\caption{Differential emission measure ($dem$) of V2129~Oph, compared to that of TW~Hya derived by \citet{BrickhouseCranmer2010}.}
\label{fig:emd_v2129oph_vs_twhya}
\end{figure}

\subsection{V2129\,Oph vs. TW Hya}

TW~Hya is considered the prototype of CTTS with high-density X-ray emitting plasma. TW~Hya is certainly not the most representative star of its class, being among the oldest CTTS. Because of its proximity, however, it is one of the best studied, especially in the X-ray band \citep{KastnerHuenemoerder2002,StelzerSchmitt2004,BrickhouseCranmer2010}.

We show in Fig.~\ref{fig:emd_v2129oph_vs_twhya} the comparison between the differential emission measure\footnote{The $dem$ is defined as $EMD/\Delta \log T$. The use of the $dem$ instead of the $EMD$ eases the comparison between models defined over temperature grids with different bin sizes since the $EM$ in a given temperature range can be simply evaluated as the area under the $dem$ curve.} ($dem$) of V2129~Oph (relative to seg.~1) and that of TW~Hya derived by \citet{BrickhouseCranmer2010}.

The X-ray emission of TW~Hya is dominated by plasma at 2.5\,MK, as evidenced by the sharp peak in the $dem$. This plasma component has very high density and is thus material heated in the accretion shock. The high temperature tail ($T\ge10$\,MK) of the $dem$ of TW~Hya indicates instead that the amount of coronal plasma is significantly lower.

The thermal structure of the plasma on V2129~Oph differs significantly from that of TW~Hya. The cool plasma on V2129~Oph has much lower $EM$ than on TW~Hya and V2129~Oph thus hosts a significantly smaller amount of high density cool plasma related to the accretion process. Conversely the corona of V2129~Oph is significantly more active than that of TW~Hya, bearing a larger amount of emission measure at high temperatures. Other than differing in absolute values, we note that V2129~Oph and TW~Hya have very different relative contributions of accretion-driven vs. coronal plasmas.

We note finally that the pole-on inclination under which TW~Hya is viewed \citep[$i\approx7\,\deg$,][]{QiHo2004} raises some questions on the observability of accretion-driven X-rays in terms of viewing geometry of the accretion shock. In fact, postulating that also TW~Hya has a hot spot latitude of $\sim60\,\deg$, then its accretion shock should be viewed at the same angle as the accretion shock of V2129~Oph during seg.~2. However accretion-driven X-rays from TW~Hya are always observed. This difference could be reconciled assuming for TW~Hya a lower latitude of the accretion shock \citep[as suggested by ][]{BatalhaBatalha2002}. Further observational constraints on the magnetospheric accretion geometry of TW~Hya are however needed to address this issue.

\section{Summary and conclusions}

Our 200\,ks {\it Chandra} observation proved that V2129~Oph is one of the X-ray brightest accreting T~Tauri stars, with $L_{\rm X}=2.7\times10^{30}\,{\rm erg\,s^{-1}}$ and $\log L_{\rm X}/L_{\rm bol} = -3.3$. The hottest components of the X-ray emitting plasma showed flaring activity typical of active coronae. Conversely the plasma at temperature of $3-4$\,MK evidenced a high density, similar to that observed in almost all CTTS with measured density, and contrasting with a coronal origin. Moreover the {\it Chandra} observation of V2129~Oph for the first time gave a hint of time variable density: during the first half of the observation the plasma at $3-4$\,MK had high density ($\log N_{\rm e} \approx 12.1$), while during the second half it showed lower density, although the statistical significance of this density difference is marginal. The fluxes of strongest emission lines produced by this cool plasma component, that probe the $EM$, are correlated with plasma density: in the first half of the observation they were significantly higher than in the second.

These results confirm that the accretion phenomenon is associated with the presence of high density plasma at a few MK, which is thus likely to be material heated in the accretion shock. They also suggest that variations of the observed density and $EM$ of this plasma component can occur over time scales of $\sim100$\,ks.

Assuming that this high density plasma is material heated in the accretion shock, these observed variations can be interpreted in terms of rotational modulation: in fact,  the viewing geometry of the accretion funnel, as determined through simultaneous optical observations, changes significantly during the {\it Chandra} observation. During the first half of the observing time the X-rays produced in the accretion shock region are observable, while during the second half of the observation the dense portions of the accretion funnel near the shock blocks the accretion-driven X-rays emitted in the direction of the observer.

\begin{acknowledgements}

We thank Fabio Reale for useful suggestions on flare analysis. SGG acknowledges the support of the Science and Technology Facilities Council [grant number ST/G006261/1].

\end{acknowledgements}

\bibliographystyle{aa} 
\bibliography{v2129oph}

\end{document}